\PassOptionsToPackage{dvipsnames}{xcolor}
\documentclass[twocolumn, switch]{article} 

\usepackage{preprint}

\usepackage{amsmath, amsthm, amssymb, amsfonts}

\usepackage[numbers,square]{natbib}
\usepackage[utf8]{inputenc}	
\usepackage[T1]{fontenc}	
\usepackage[colorlinks = true,
            linkcolor = purple,
            urlcolor  = blue,
            citecolor = cyan,
            anchorcolor = black]{hyperref}	
\usepackage{nicefrac}		
\usepackage{microtype}		
\usepackage{lineno}		
\usepackage{float}			

\usepackage{newfloat}
\DeclareFloatingEnvironment[name={Supplementary Figure}]{suppfigure}
\usepackage{sidecap}
\sidecaptionvpos{figure}{c}

\usepackage{titlesec}
\titlespacing\section{0pt}{12pt plus 3pt minus 3pt}{1pt plus 1pt minus 1pt}
\titlespacing\subsection{0pt}{10pt plus 3pt minus 3pt}{1pt plus 1pt minus 1pt}
\titlespacing\subsubsection{0pt}{8pt plus 3pt minus 3pt}{1pt plus 1pt minus 1pt}

\usepackage[toc,page]{appendix} 
\usepackage{subcaption} 
\usepackage{booktabs} 
\usepackage{dsfont} 
\usepackage{amssymb}
\usepackage{pifont}
\usepackage{stmaryrd} 
\usepackage{makecell} 
\usepackage{graphicx,multirow} 
\usepackage{colortbl} 
\usepackage{amsmath}
\usepackage{bm} 
\usepackage{url}  
\usepackage{siunitx} 
\usepackage{caption} 
\usepackage{float} 
\usepackage{threeparttable}
\usepackage{tabularx}

\definecolor{goldyellow}{RGB}{252, 214, 0}
\definecolor{redd}{RGB}{199, 53, 0}
\definecolor{greenn}{RGB}{213,232,212}
\definecolor{bluee}{RGB}{218,232,252}

\definecolor{Gray}{gray}{0.96}
\newcolumntype{g}{>{\columncolor{Gray}}c}
\usepackage{etoolbox}
\usepackage{tablefootnote}

\title{CellViT: Vision Transformers for Precise Cell Segmentation and Classification}

\usepackage{background}
\backgroundsetup{
    angle=0,  
    scale=0.75,
    placement=bottom,
    color=Black, 
    vshift=20pt,
    contents={\ifnum\value{page}=1 *Corresponding author: \tt{fabian.hoerst@uk-essen.de} \else  \fi}
}

\usepackage{authblk}

\author[1,2\thanks{}]{Fabian Hörst}

\author[1,2]{Moritz Rempe}
\author[1,2]{Lukas Heine}
\author[1,3]{Constantin Seibold}
\author[1,4]{Julius Keyl}
\author[1,5]{Giulia Baldini}
\author[6,7]{Selma Ugurel}
\author[8,9,10,11]{Jens Siveke}
\author[12,13]{Barbara Grünwald}
\author[1,2]{Jan Egger}
\author[1,2,7,14]{Jens Kleesiek}

\affil[1]{\small Institute for AI in Medicine (IKIM), University Hospital Essen (AöR), Essen, Germany}
\affil[2]{Cancer Research Center Cologne Essen (CCCE), West German Cancer Center Essen, University Hospital Essen (AöR), Essen, Germany}
\affil[3]{Clinic for Nuclear Medicine, University Hospital Essen (AöR), Essen, Germany}
\affil[4]{Institute of Pathology, University Hospital Essen (AöR), Essen, Germany}
\affil[5]{Institute of Interventional and Diagnostic Radiology and Neuroradiology, University Hospital Essen (AöR), Essen, Germany}
\affil[6]{Department of Dermatology, University Hospital Essen (AöR), Essen, Germany}
\affil[7]{German Cancer Consortium (DKTK, Partner site Essen), Heidelberg, Germany}
\affil[8]{Bridge Institute of Experimental Tumor Therapy, West German Cancer Center Essen, University Hospital Essen (AöR), Essen, Germany }
\affil[9]{Division of Solid Tumor Translational Oncology, German Cancer Consortium (DKTK, Partner site Essen) and German Cancer Research Center (DKFZ), Heidelberg, German}
\affil[10]{West German Cancer Center, Department of Medical Oncology, University Hospital Essen (AöR), Essen, Germany}
\affil[11]{Medical Faculty, University Duisburg-Essen, Essen, Germany}
\affil[12]{Department of Urology, West German Cancer Center, University Hospital Essen (AöR), Germany}
\affil[13]{Princess Margaret Cancer Centre, Toronto, Ontario, Canada}
\affil[14]{Department of Physics, TU Dortmund University, Dortmund, Germany}

\normalsize

\begin{document}
\twocolumn[ 
  \begin{@twocolumnfalse} 
  
\maketitle
\vspace{-2mm}
\begin{abstract}
    Nuclei detection and segmentation in hematoxylin and eosin-stained (H\&E) tissue images are important clinical tasks and crucial for a wide range of applications. However, it is a challenging task due to  nuclei variances in staining and size, overlapping boundaries, and nuclei clustering. While convolutional neural networks have been extensively used for this task, we explore the potential of Transformer-based networks in this domain. Therefore, we introduce a new method for automated instance segmentation of cell nuclei in digitized tissue samples using a deep learning architecture based on Vision Transformer called CellViT. CellViT is trained and evaluated on the PanNuke dataset, which is one of the most challenging nuclei instance segmentation datasets, consisting of nearly 200,000 annotated Nuclei into 5 clinically important classes in 19 tissue types.
    We demonstrate the superiority of large-scale in-domain and out-of-domain pre-trained Vision Transformers by leveraging the recently published \textit{Segment Anything Model} and a ViT-encoder pre-trained on 104 million histological image patches - achieving state-of-the-art nuclei detection and instance segmentation performance on the PanNuke dataset with a mean panoptic quality of $0.50$ and an $F_1$-detection score of $0.83$. The code is publicly available at \url{https://github.com/TIO-IKIM/CellViT}. 
\end{abstract}

\keywords{Cell Segmentation \and Digital Pathology \and Deep Learning \and Computer Vision \and Vision Transformer \and Segment Anything}
\vspace{0.35cm}

  \end{@twocolumnfalse} 
]

\section{Introduction}
\label{sec:introduction}
Cancer is a severe disease burden worldwide, with millions of new cases yearly and ranking as the second leading cause of death after cardiovascular diseases \citep{Cancer2022}. Despite novel and powerful non-invasive radiological imaging modalities, collecting tissue samples and evaluating them with a microscope remains a standard procedure for diagnostic evaluation. A pathologist can draw conclusions about potential therapeutic approaches or use them as a starting point for further investigations by identifying abnormalities within the tissue. One crucial component is the analysis of the cells and their distribution within the tissue, such as detecting tumor-infiltrating lymphocytes \citep{TIL} or inflammatory cells in the tumor microenvironment \citep{tme_analysis, tme_gruenwald}. However, large-scale analysis on the cell level is time-consuming and suffers from a high intra- and inter-observer variability. \\
Due to the development of high-throughput scanners for pathology, it is now possible to create digitized tissue samples (whole-slide images, WSI), enabling the application of computer vision (CV) algorithms. CV facilitates automated slide analysis, for example, to create tissue segmentation \citep{vv}, detect tumors \citep{clam}, evaluate therapy response \citep{memori}, and the computer-aided detection and segmentation of cells \citep{Hovernet, tsfd}. In addition to the clinical applications mentioned above, cell instance segmentation can be leveraged for downstream deep learning tasks, as each WSI contains numerous nuclei of diverse types, fostering systematic analysis and predictive insights \citep{GRAHAMOneModell}. \citet{Sirinukunwattana2018} showed that cell analysis supports the creation of high-level tissue segmentation based on cell composition. \citet{cell_classifier} used hand-crafted features extracted from cells to detect tumor regions in a slide. \\
Existing algorithms for analyzing WSI \citep{clam, Campanella2019, memori} are often based on Convolutional Neural Networks (CNNs) used as feature extractors for image regions. The algorithms, despite achieving clinical-grade performance \citep{Campanella2019}, face limitations in interpretability, which in turn poses challenges in defining novel human-interpretable biomarkers. However, accurate cell analysis within these slides presents an opportunity to construct explainable pipelines, incorporating human-interpretable features effectively in downstream tasks \citep{GRAHAMOneModell, graham2021conic}. Nevertheless, since subtask WSI analysis models \citep{clam, Campanella2019, memori} rely on abstract entity embeddings, features must be extracted from the detected cells. One approach is to generate hand-crafted features, such as morphological attributes, from the segmentation \citep{cell_features}. In the radiology setting, this is referred to as Radiomics \citep{radiomics}. Alternatively, employing a CNN on image sections of single cells can derive deep learning features. While hand-crafted features may have limited performance, using CNNs for each cell is computationally complex. 
Thus, the need for automated and reliable detection and segmentation of cells in conjunction with cell-feature extraction in WSI is evident. \\
\begin{figure}[!t]
    \centering
    \includegraphics[width=\columnwidth]{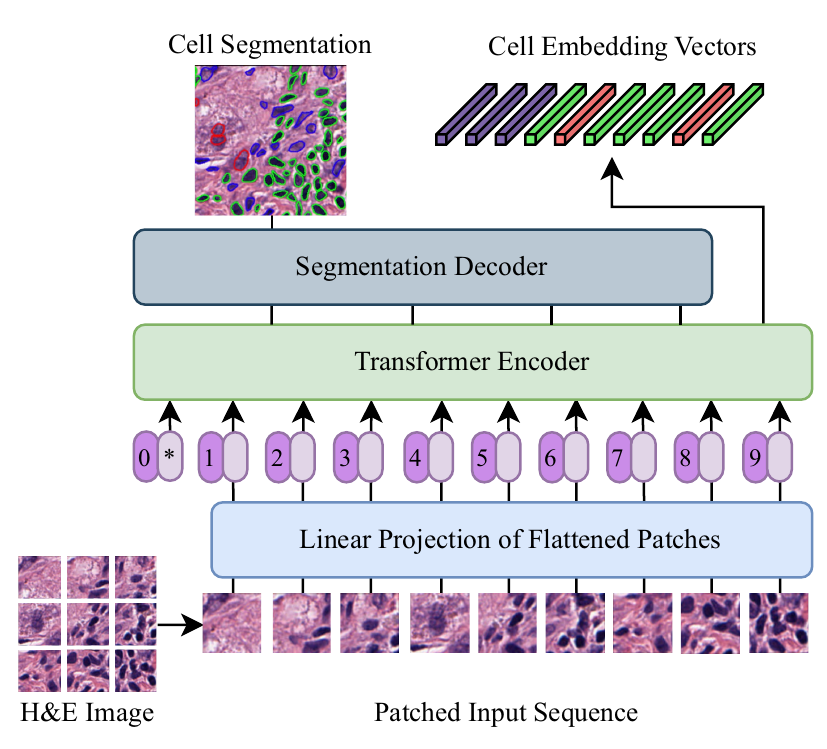}
    \caption{Network structure of CellViT. An input image is transformed into a sequence of tokens (flattened input sections). By using skip connections at multiple encoder depth levels and a dedicated upsampling decoder network, precise nuclei instance segmentations are derived. Nuclei embeddings are extracted from the Transformer encoder.} 
    \label{fig:network_high_level}
    \vspace{-3mm}
\end{figure} 
We developed a novel deep learning architecture based on Vision Transformer for automated instance segmentation of cell nuclei in digitized tissue samples (\textbf{CellViT}). Our approach eliminates the need for additional computational effort for deriving cell features via parallel feature extraction during runtime. The \mbox{CellViT} model proves to be highly effective in collecting nuclei information within patient cohorts and could serve as a reliable nucleus feature extractor for downstream algorithms. Our solution demonstrates exceptional performance on the PanNuke \citep{pannuke} dataset by leveraging transfer learning and pre-trained models \citep{vit256, SAM}. The PanNuke dataset contains $189{,}744$ segmented nuclei and includes 19 different types of tissues. Among these tissues, there are five clinically important nuclei classes: Neoplastic, inflammatory, epithelial, dead, and connective/soft cells.
In addition to the high number of tissue classes and nuclei types, the dataset is highly imbalanced, creating additional complexity. Besides class imbalance, segmenting cell nuclei itself is a difficult task. The cell nuclei may overlap, have a high level of heterogeneity and inter- or intra-instance variability in shape, size, and staining \citep{tsfd}. Sophisticated training methods such as transfer learning, data augmentation, and specific training sampling strategies next to postprocessing algorithms are necessary to achieve satisfactory results. \\

The proposed network architecture is based on a U-Net-shaped encoder-decoder architecture similar to HoVer-Net \citep{Hovernet}, one of the leading models for nuclei segmentation. Notably, we replace the traditional CNN-based encoder network with a Vision Transformer, inspired by the UNETR architecture \citep{unetr}. This approach is depicted in Figure \ref{fig:network_high_level}. Vision Transformers are token-based neural networks that use the attention mechanism to capture both local and global context information. This ability enables ViTs to understand relationships among all cells in an image, leveraging long-range dependencies and substantially improving their segmentation. Moreover, when using the common token size of $16~\text{pixels}~(\text{px})$ and pixel-resolutions such as $0.25~\si{\micro\meter \per px}$ (commonly $\times 40$ magnification) or $0.50~\si{\micro\meter \per px}$ (commonly $\times 20$ magnification) of the images, the token size of ViTs is approximately equivalent to that of a cell, enabling a direct association between a detected cell and its corresponding token embedding from the ViT encoder. As a result, we directly obtain a localizable feature vector during our cell detection that we can extract simultaneously within one forward pass, unlike CNN networks.
\\
Given the limited amount of available data in the medical domain, pre-trained models are an essential requirement as ViTs have increased data requirements compared to CNNs. \citet{vit256} recently published a ViT pre-trained on 104 million histological images ($\textit{ViT}_\textit{256}$). Their network outperformed current state-of-the-art (SOTA) cancer subtyping and survival prediction methods.
Another important contribution is the \textit{Segment Anything Model (SAM)}, proposed by \citet{SAM}.
They developed a generic segmentation network for various image types, whose zero-shot performance is almost equivalent to many supervised trained networks. In our work, we compare the performance of pre-trained $\text{ViT}_{256}$ \citep{vit256} and SAM \citep{SAM} models as building blocks of our architecture for nuclei segmentation and classification. We demonstrate superior performance over existing nuclei instance segmentation models. 
We summarize our contributions as follows:
\begin{enumerate}
    \item We present a novel U-Net-shaped encoder-decoder network for nuclei instance segmentation, levering Vision Transformers as encoder networks. Our approach surpasses existing methods for nuclei detection by a substantial margin and achieves competitive segmentation results with other state-of-the-art methods on the PanNuke dataset. We demonstrate the generalizability of CellViT by applying it to the MoNuSeg dataset without finetuning.
    \item We are the first to employ Vision Transformer networks for nuclei instance segmentation on the PanNuke dataset, demonstrating their effectiveness in this domain. The proposed approach combines pre-trained ViT encoders with a decoder network connected by skip connections.  
    \item We provide a framework that enables fast inference results applied on Gigapixel WSI by using a large inference patch size of $1024 \times 1024~\text{px}$ in contrast to conventional $256~\text{px}$-sized patches. Compared to HoVer-Net, our inference pipeline runs $1.85$ times faster.
\end{enumerate}

\section{Related Work}
\vspace{3mm}
\label{sec:related_works}
\subsection{Instance Segmentation of Nuclei} 
\vspace{3mm}
\label{subsec:semantic_segmentation_of_nuclei}
Numerous methods have been developed to solve the challenging task of cell nuclei instance segmentation in WSIs. Previous works have explored diverse approaches, ranging from traditional image processing techniques to deep learning (DL) methods. Commonly used image processing techniques involve the design and extraction of domain-specific features. These features encompass characteristics such as intensity, texture, shape, and morphological properties of the nuclei. The primary challenge is separating overlapping nuclei, and different techniques have been devised to do this \citep{watershed_2006, Malpica1998_watershed, tareef_2018_watershed, cheng_2009, veta_2013, ali_2012, Wienert2012, Liao2016}.
For instance, the works of \citet{cheng_2009}, \citet{veta_2013}, and \citet{ali_2012} rely on a predefined nuclei geometry and the watershed algorithm to separate clustered nuclei, while \citet{Wienert2012} used morphological operations without watershed and \citet{Liao2016} utilized eclipse-fitting for cluster separation. A common drawback of these techniques is their dependency on hand-crafted features, which require expert-level domain knowledge, have limited representative power, and are sensitive to hyperparameter selection \citep{Hovernet, cpp_net}. The complexity of extracting meaningful features increases when cell nuclei classification is added to the segmentation task. Consequently, their performance is insufficient for our needs to classify and segment nuclei in various tissue types \citep{cpp_net}.\\

To overcome the limitations of traditional image processing techniques, DL has emerged as a powerful approach for nuclei instance segmentation. An inherent advantage of DL networks is their automatic extraction of relevant features for the given task, surpassing the need for expert-level domain knowledge to generate hand-crafted features. DL algorithms, particularly convolutional neural networks (CNNs) \citep{image_segmentation_using_dl, dl_healthcare}, have shown remarkable success in various computer vision tasks \citep{LeCun2015}. Especially the invention of the U-Net architecture by \citet{ronneberger2015unet} has significantly impacted medical image analysis by enabling accurate and efficient segmentation of complex structures, contributing to advancements in various medical domains such as radiology \citep{nnunet, radiology_unet} and digital pathology \citep{medical_unet}. It consists of a U-shaped encoder-decoder structure with skip connections at multiple network depths to preserve fine-grained details in the decoder. However, the original U-Net implementation is not able to separate clustered nuclei \citep{Hovernet}. Therefore, specialized network architectures are necessary to separate clustered and overlapping cell nuclei. In the current literature, DL algorithms for nuclei instance segmentation are further divided into two-stage and one-stage methods \citep{tsfd}. \\
Two-stage methods incorporate a cell detection network in the first stage to localize cell nuclei within an image, generating bounding box predictions of nuclei. These detected nuclei are then passed on to a subsequent segmentation stage to retrieve a fine-grained nucleus segmentation.
Mask-RCNN \citep{mask_rcnn} is one of the leading two-stage models built on top of the object detection model Fast-RCNN \citep{fast_rcnn}. \citet{mask_rcnn_adaption_2} utilized Mask-RCNN networks for nuclei instance segmentation. Based on the proposed nuclei detections in the first stage, the model incorporates a segmentation branch for the fine-grained nucleus segmentations in the second stage. A rectangular image section of the detected nuclei is used as input for the segmentation stage, which causes the problem that overlapping neighboring nuclei may be segmented as well and need to be cleaned up by an additional postprocessing algorithm. Another two-stage method for nuclei segmentation is BRP-Net \citep{brp_net}, which creates nuclei proposals in the first place, then refines the boundary, and finally creates a segmentation out of this. However, this network structure is computationally complex and not designed for end-to-end training due to three independent stages. Additionally, the network requires a considerable time of 12 minutes to segment a $1360\times1024~\text{px}$ image, making its practical application nearly impossible \citep{brp_net}. While two-stage systems offer advantages in localizing cells and improving individual nucleus detection, they often require additional postprocessing for segmentation and suffer from time and computational complexity. \\
In comparison, one-stage methods combine a single DL network with postprocessing operations. Micro-Net \citep{micro_net} extends the U-Net by using multiple resolution input images to be invariant against nuclei of varying sizes. The DIST model by \citet{dist} adds an additional decoder branch next to the segmentation branch to detect nuclei markers for a watershed postprocessing algorithm. For this, they predict distance maps from the nucleus boundary to the center of mass of the nuclei. Distance maps are regression maps indicating the distance of a pixel to a reference point, e.g., from a nuclei pixel to the center of mass. HoVer-Net \citep{Hovernet}, one of the current SOTA methods for automatic nuclei instance segmentation, uses horizontal and vertical distances of nuclei pixels to their center of mass and separates the nuclei by using the gradient of the horizontal and vertical distance maps as an input to an edge detection filter (Sobel operator). The models STARDIST \citep{stardist, stardist2} and its extension CPP-Net \citep{cpp_net} generate polygons defining the nuclei boundaries over a set of predicted distances. For this, STARDIST utilizes a star-convex polygon representation to approximate the shape of nuclei. Whereas in STARDIST, the polygons are derived just by features of the centroid pixel, CPP-Net uses context information from sampled points within a nucleus and proposes a shape-aware perceptual loss to constrain the polygon shape. STARDIST demonstrates comparable segmentation performance to HoVer-Net, while CPP-Net exhibits slightly superior results. \\

In contrast, boundary-based methods such as DCAN \citep{dcan} and TSFD-Net \citep{tsfd} adopt a different approach, where instead of using distance maps, watershed markers, or polygon predictions, they directly predict the nuclear contour using a prediction map. While DCAN is based on the U-Net architecture, TSFD-Net utilizes a Feature Pyramid Network (FPN) \citep{fpn} to leverage multiple scales of features. Additionally, the authors of TSFD-Net introduce a tissue-classifier branch to learn tissue-specific features and guide the learning process. To address the class imbalance across nuclei and tissue types, they employ the focal loss \citep{focal_loss} for the tissue detection branch, a modified cross-entropy loss with dynamic scaling, and the Focal Tversky loss \citep{focal_tversky_loss} for the segmentation branch, which enlarges the contribution of challenging regions. While TSFD-Net shows promising results, its comparability to other methods is limited due to the lack of a standardized evaluation procedure.

\subsection{Vision Transformer}
\label{subsec:vision_transformer}
All promising DL models \citep{mask_rcnn, brp_net,micro_net,dist,stardist,Hovernet,cpp_net,dcan,tsfd} for nuclei instance segmentation mentioned previously are based on CNNs. Even though CNN models have demonstrated their effectiveness in image processing, they are bound to local receptive fields and may struggle to capture spatial long-range relationships \citep{vv}. Inspired by the Transformer architecture in NLP \citep{attention_is_all_you_need}, Vision Transformers \citep{vit_orig} have recently emerged as an alternative to CNNs for CV \citep{dino}. Their architecture is based on the self-attention mechanism \citep{attention_is_all_you_need}, allowing the model to attend to any region within an image to capture long-range dependencies. Unlike CNNs, they are also not bound to fixed input sizes and can process images of arbitrary sizes depending on computational capacity. Vision Transformers have shown promising results not only in image classification \citep{vit_orig,dino,do_vits_see_like_cnns}, but also in other vision tasks such as object detection \citep{vit_yolo} and semantic segmentation \citep{unetr,vv}. \\

\vspace{-2mm}
\paragraph{Vision Transformers for Instance Segmentation}
In recent years, various ideas to use the Transformer architecture for instance segmentation have been developed \citep{transunet, li2021medical, unetr, swin_unetr, segformer, rethinking_semantic_seg}. Primarily, these methods integrate Transformer models into encoder-decoder architectures by exchanging or extending the encoder network of existing U-Net-based solutions. \citet{transunet} used a Transformer in their TransUNet network to encode tokenized patches from a CNN feature map as the input sequence to derive global context within the CNN network. \citet{li2021medical} applied a squeeze-and-expansion Transformer as a variant of the original Vision Transformer by \citet{vit_orig} for medical segmentation. The Segformer model by \citet{segformer} incorporates an adapted Transformer as an image encoder connected to a lightweight MLP decoder segmentation head. In contrast to these methods, the SETR model \citep{rethinking_semantic_seg}, used the original ViT as encoder and a fully convolution network as decoder, both connected without intermediate skip connections. Building upon these advancements, the UNETR model \citep{unetr} combined a standard ViT connected to a U-Net-like decoder with skip connections, outperforming TransUNet and the SETR model on three medical image segmentation datasets. The integration of the original ViT implementation without adaptions into the powerful U-Net framework allows the use of pre-trained ViT-networks, which is an important property exploited in our work. \\

\vspace{-2mm}
\paragraph{Large-scale Pre-Training}
\label{subsec:large_scale_pretraining}
Pre-training a Vision Transformer on a large amount of data serves as a crucial step to initialize the model's parameters with meaningful representations. \citet{vit_orig} demonstrated that ViTs require a larger amount of data compared to CNNs to learn meaningful representations.  This is attributed to the inductive biases of the receptive fields of CNNs that are useful for smaller datasets. In contrast, ViTs need to learn relevant patterns, but when provided with sufficiently large datasets, these patterns are more meaningful \citep{do_vits_see_like_cnns}. In the medical domain, where annotated data is often limited, pre-trained ViT-based networks become even more critical. By utilizing self-supervised pre-training approaches \citep{simclr, moco, swav, byol, simsiam, dino}, available unlabelled data can be facilitated effectively to initialize network weights before finetuning the network on the target domain. One popular self-supervised pre-training approach, specifically adapted for Vision Transformers, is DINO (knowledge distillation with no labels) \citep{dino}. Vision Transformers trained with this method contain features that explicitly include information about the semantic segmentation of images, which does not emerge as clearly with CNNs \citep{dino}. \\ 
In the histopathological domain, \citet{vit256} developed a hierarchical network for slide-level representation by stacking multiple ViT blocks. Their approach involves a three-stage hierarchical architecture performing a bottom-up aggregation, with each stage pre-trained independently with DINO. The first stage focuses on processing $16 \times 16~\text{px}$-sized visual tokens out of $256 \times 256~\text{px}$ patches to create a local cell-cluster token. This first stage ViT, which we refer to as $\textbf{ViT}_\mathbf{256}$ (ViT-Small, 
 $21.7~\text{M}$ parameter), is particularly relevant for semantic segmentation. The authors pre-trained the $\text{ViT}_{256}$ on 104 million $256 \times 256~\text{px}$-sized histological image patches from The Cancer Genome Atlas (TCGA) and made the network weights publicly available. It was demonstrated that the $\text{ViT}_{256}$ network successfully learned visual concepts specific to histopathological tissue images, including fine-grained cell locations, stroma, and tumor regions, making the model a powerful pre-trained backbone network for histological image analysis. \\
As for the "natural image"-domain, \citet{SAM} recently published a promptable open-source segmentation model as a "foundation model" \citep{foundation_models} for semantic segmentation, also known as \textbf{Segment Anything (SAM)}. The SAM framework comprises an image encoder (ViT) and a lightweight mask decoder network. The final backbone (ViT-H) of SAM was trained supervised on $1.1$ billion segmentation masks from $11$ million images. A three-stage data engine consisting of assisted manual, semi-automatic, and automatic mask generation acquired this extensively annotated dataset. Pre-trained weights for three different ViT-scales (ViT-Base with $86~\text{M}$ parameter, denoted as SAM-B, ViT-Large with $307~\text{M}$ parameter, denoted as SAM-L, and ViT-Huge with $632~\text{M}$ parameter, denoted as SAM-H) are publicly available.

\section{Methods}
\label{sec:methods}
Our architecture is inspired by the UNETR model \citep{unetr} for 3D volumetric images, but we adapt its architecture for processing 2D images as shown in Fig. \ref{fig:network_detail}. Unlike traditional segmentation networks that employ a single decoder branch for computing the segmentation map, our network employs three distinct multi-task output branches inspired by the approach of HoVer-Net \citep{Hovernet}. The first branch predicts the binary segmentation map of all nuclei (nuclei prediction, NP), capturing their boundaries and shapes. The second branch generates horizontal and vertical distance maps (horizontal-vertical prediction, HV), providing crucial spatial information for precise localization and delineation. Lastly, the third branch predicts the nuclei type map (NT), enabling the classification of different nucleus types.
In summary, our network has the following multi-task branches for instance segmentation:
\begin{itemize}
    \item \makebox[1.6cm]{NP-branch:\hfill} Predicts binary nuclei map 
    \item \makebox[1.6cm]{HV-branch:\hfill} Predicts the horizontal and vertical distances of nuclear pixels to their center of mass, normalized between -1 and 1 for each nuclei
    \item \makebox[1.6cm]{NT-branch:\hfill} Predicts the nuclei types as instance segmentation maps
\end{itemize}

To integrate these outputs, we utilize additional postprocessing steps. These steps involve merging the information from the different branches, separating overlapping nuclei to ensure accurate individual segmentation, and determining the nuclei class based on the nuclei type map. \\
In our exepriments, we also evaluated the effectiveness of the STARDIST decoder method and its extension, CPP-Net. We integrate their techniques into the proposed UNETR-HoVer-Net architecture with modifications. Instead of the NP-branch, an object probability branch \textit{PD} is used to predict whether a pixel belongs to an object by predicting the Euclidean distance to the nearest background pixel. The HV-branch is replaced by a branch \textit{RD} to predict the radial distances of an object pixel to the boundary of the nucleus (star-convex representation) \citep{stardist}. The NT-branch remains unchanged. For the CPP-Net decoder, an additional refinement step is added for the radial distances \citep{cpp_net}.

\subsection{Network Structure}
\label{subsec:network_structures}
\begin{figure*}[!t]
	\centering
	\includegraphics[width=\textwidth]{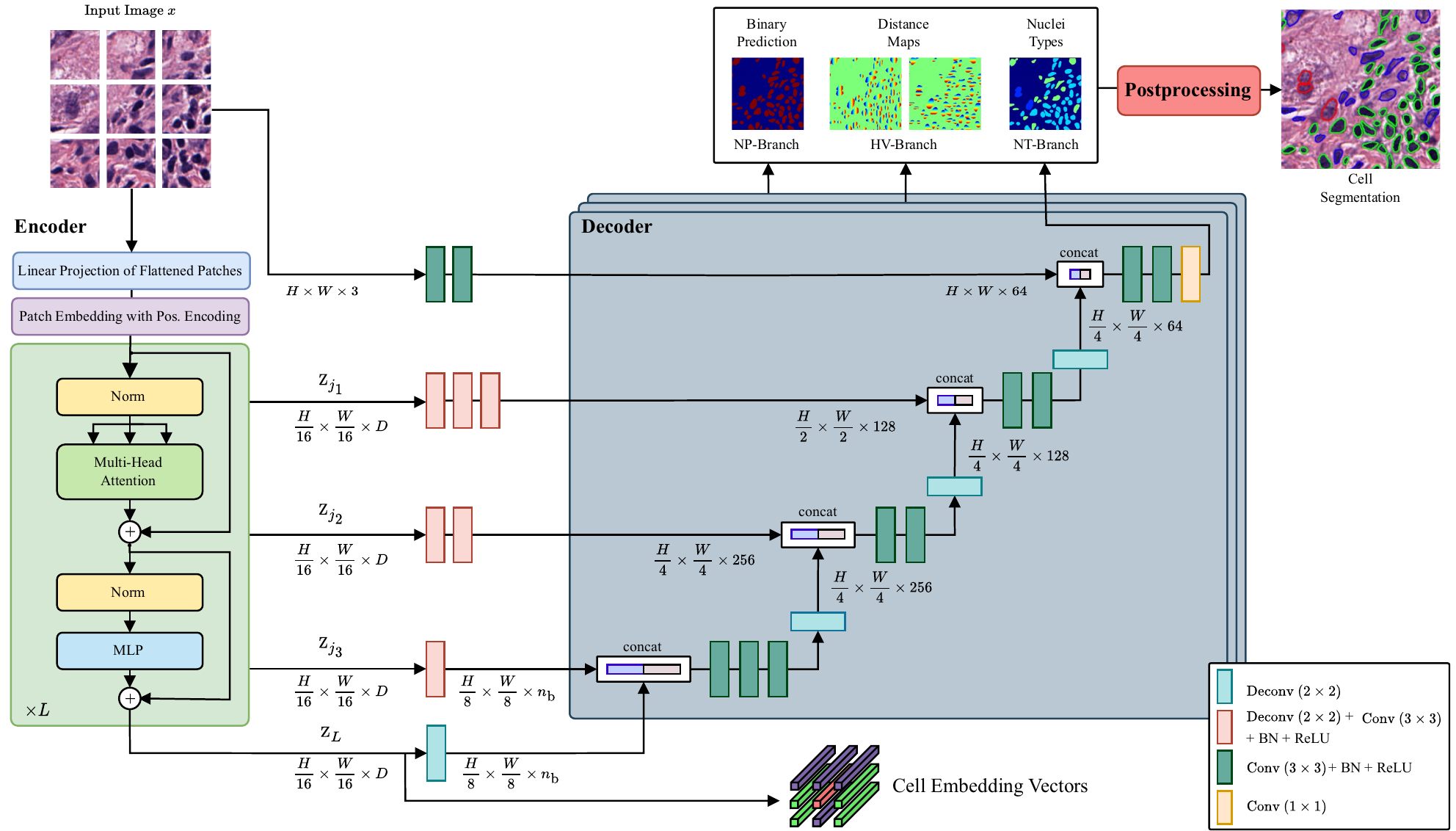}
	\caption{Network structure of our proposed CellViT-network consisting of a ViT encoder connected to multiple decoders via skip connections. Postprocessing is used to separate overlapping nuclei and perform nuclei type classification. For visualization purposes, the tissue classification branch is not illustrated. As encoder networks, we used the pre-trained $\text{ViT}_{256}$ and SAM models.} 
	\label{fig:network_detail}
\end{figure*}
In our network, we integrate a Vision Transformer as an image encoder that is connected to an upsampling decoder network via skip connections. This architecture allows us to leverage the strengths of a Vision Transformer as an image encoder for instance segmentation without losing fine-grained information. Even though many other adaptations of the U-Net structure for Vision Transformers have been proposed (e.g., SwinUNETR \citep{swin_unetr}), it was important for us to choose a network structure that incorporates the original ViT structure by \citet{vit_orig}  without modifications such that we can make use of the large-scale pre-trained ViTs, namely $\text{ViT}_{256}$ and SAM. \\
As in NLP \citep{attention_is_all_you_need}, Vision Transformers take as input a 1D sequence of tokens embeddings \citep{vit_orig, attention_is_all_you_need}. Therefore we need to divide an input image $\bm{x} \in \mathbb{R}^{H \times W \times C}$ with height $H$, width $W$ and $C$ input channels into a sequence of flattened tokens $\bm{x}_\text{p} \in \mathbb{R}^{N \times \left(P^2\cdot C \right )}$. Each token is a squared image section with the dimension $P \times P$. The number of tokens $N$ can be calculated via $N=HW\slash P^2$, which is the effective input sequence length \citep{unetr}. Accordingly, a linear projection layer $\bm{E} \in \mathbb{R}^{N \times D}$ is used to map the flattened tokens $\bm{x}_\text{p}$ into a $D$-dimensional latent space. The latent vector size $D$ remains constant through all of the Transformer layers.
In contrast to the UNETR-network, we incorporate a learnable class token $\bm{x}_\text{class}$ \citep{vit_orig}, which we can use for classification tasks and append it to the token sequence. \\
Unlike CNNs, which inherently capture spatial relationships through their local receptive fields, Transformers are permutation invariant and, therefore, cannot capture spatial relationships. Thus, a learnable 1D positional embedding $\bm{E}_\text{pos} \in \mathbb{R}^{\left(N+1 \right ) \times D}$ is added to the projected token embeddings to preserve spatial context \citep{unetr}. In summary, the final input sequence $\bm{z}_0$ for the Transformer encoder is: 
\begin{align}
    \bm{z}_0 = \left [\bm{x}_\text{class}; \bm{x}^1_\text{p}\bm{E}; \bm{x}^2_\text{p}\bm{E}; \hdots ; \bm{x}^N_\text{p}\bm{E}\right] + \bm{E}_\text{pos}. 
\end{align}
The Transformer encoder comprises alternating layers of multiheaded
self-attention (MHA) \citep{vit_orig} and multilayer perceptrons (MLP), assembled in one Transformer block. A ViT is composed of several stacked Transformer blocks such that the latent tokens $\bm{z}_i$ are calculated by 
\begin{align}
    \bm{z}^{'}_i &= \text{MHA}(\text{Norm}(z_{i-1})) + \bm{z}_{i-1},~i=1 \hdots L \\
    \bm{z}_i &= \text{MLP}(\text{Norm}(z_{i-1})) + \bm{z}_{i-1},~i=1 \hdots L,
\end{align}
with $L$ denoting the number of Transformer blocks, $\text{Norm}(\cdot)$ denoting layer normalization, and $i$ is the intermediate block identifier \citep{unetr}. 
Inspired by the U-Net and UNETR architectures, we add skip connections to leverage information at multiple encoder depths in the decoder. In total, we use five skip connections. The first skip connection takes $\bm{x}$ as input and processes it by two convolution layers ($3 \times 3$ kernel size) with batch-normalization and ReLU activation functions. For the remaining four skip connections, the intermediate and bottleneck latent tokens $\bm{z}_j,~j \in \left \{ \frac{L}{4}, \frac{2L}{4}, \frac{3L}{4}, L\right \}$ are extracted without the class token and reshaped to a 2D tensor 
$\bm{Z}_j \in \mathbb{R}^{\frac{H}{P} \times \frac{W}{P}\times D}$. This is only valid if $4\mid L$ holds, which is commonly satisfied for common ViT implementations \citep{vit_orig, vit256, SAM}. Each of the feature maps $\bm{Z}_j$ is transformed by a combination of deconvolutional layers that increase the resolution in both directions by a factor of two and convolutions to adjust the latent dimension. Subsequently, the transformed feature maps are successively processed in each decoder, beginning with $\bm{Z}_L$, and fused with the corresponding skip connection at each stage. This iterative fusion ensures the effective incorporation of multi-scale information, enhancing the overall performance of the decoder. Our network is designed in such a way that the output resolution of the segmentation results exactly matches the input image resolution. \\
As denoted in Fig. \ref{fig:network_detail}, our three segmentation branches (NP, HV, NT) share the same image encoder with the same skip connections and their transformations. The only difference lies in the isolated upsampling pathways of the decoders specific to each branch. \\
To leverage the additional tissue type information available in the PanNuke dataset, we introduce a tissue classification branch (TC) to guide the learning process of the encoder. For this, we use the class token $z_{L,\text{class}}$ as input to a linear layer with softmax activation function to predict the tissue class. 

\subsection{Target and Losses}
\label{subsec:target_and_losses}
For faster training and better convergence of the network, we employ a combination of different loss functions for each network branch. The total loss is 
\begin{align}
    \label{eq:loss_total}
    \mathcal{L}_\text{total} = \mathcal{L}_\text{NP} + \mathcal{L}_\text{HV} + \mathcal{L}_\text{NT} + \mathcal{L}_\text{TC}
\end{align}
where $\mathcal{L}_\text{NP}$ denotes the loss for the NP-branch, $\mathcal{L}_\text{HV}$ the loss for the HV-branch, $\mathcal{L}_\text{NT}$ the loss for the NT-branch, and  $\mathcal{L}_\text{TC}$ the loss for the TC-branch. 
Overall, the individual branch losses are composed of the following weighted loss functions:
\begin{align}
\label{eq:branch_losses}
\begin{split}
    \mathcal{L}_\text{NP} &= \lambda_{\text{NP}_\text{FT}}\mathcal{L}_\text{FT} + \lambda_{\text{NP}_\text{DICE}}\mathcal{L}_\text{DICE}  \\ 
    \mathcal{L}_\text{HV} &= \lambda_{\text{HV}_\text{MSE}}\mathcal{L}_\text{MSE} + \lambda_{\text{HV}_\text{MSGE}}\mathcal{L}_\text{MSGE}  \\
    \mathcal{L}_\text{NT} &= \lambda_{\text{NT}_\text{FT}}\mathcal{L}_\text{FT} + \lambda_{\text{NT}_\text{DICE}}\mathcal{L}_\text{DICE} + \lambda_{\text{NT}_\text{BCE}}\mathcal{L}_\text{BCE}\\
    \mathcal{L}_\text{TC} &= \lambda_{\text{TC}_\text{CE}}\mathcal{L}_\text{CE} 
\end{split}
\end{align}
with the individual segmentation losses
\small\begin{align}
    \mathcal{L}_\text{BCE} &= -\frac{1}{n} \sum_{i=1}^{N_\text{px}} \sum_{c=1}^{C} y_{i,c} \log \hat{y}_{i c} \label{eq:bce}\\
    \mathcal{L}_\text{DICE} &= 1-\frac{2 \times \sum_{i=1}^{N_\text{px}} y_{i c} \hat{y}_{i c}+\varepsilon}{\sum_{i=1}^{N_\text{px}} y_{i c} + \sum_{i=1}^{N_\text{px}} \hat{y}_{i c} + \varepsilon} \label{eq:dice}\\
    \mathcal{L}_\text{FT} &= \sum_{c=1}^C \left(1-\frac{\sum_{i=1}^{N_\text{px}} y_{i c} \hat{y}_{i c}+\varepsilon}{\sum_{i=1}^{N_\text{px}} y_{i c} \hat{y}_{i c}+\alpha_{\text{FT}} \sum_{i=1}^{N_\text{px}} y_{i c} \hat{y}_{i c}+\beta_{\text{FT}} \sum_{i=1}^{N_\text{px}} y_{i c} \hat{y}_{i c}}\right)^{\frac{1}{\gamma_{\text{FT}}}} \label{eq:focal_tversky}
\end{align}\normalsize
\vspace{-1mm} and the cross-entropy as tissue classification loss
\begin{align*}
    \mathcal{\mathcal{L}_\text{CE}} &= -\sum_{c_\text{T}=1}^{C_\text{T}} y_{c_\text{T}} \log \hat{y}_{c_\text{T}},~~C_\text{T}=19, 
\end{align*}
with the contribution of each branch loss \eqref{eq:branch_losses} to the total loss \eqref{eq:loss_total} controlled by the $i-\text{th}$ hyperparameters $\lambda_i$. $\mathcal{L}_\text{MSE}$ denotes the mean squared error of the horizontal and vertical distance maps and $\mathcal{L}_\text{MSGE}$ the mean squared error of the gradients of the horizontal and vertical distance maps, each summarized for both directions separately. In the segmentation losses \eqref{eq:bce}-\eqref{eq:focal_tversky}, $y_{ic}$ is the ground-truth and $\hat{y}_{i c}$ the prediction probability of the $i$th pixel belonging to the class $c$, $C$ the total number of nuclei classes, $N_\text{px}$ the total amount of pixels, $\varepsilon$ a smoothness factor and  $\alpha_{\text{FT}}, \beta_{\text{FT}}$ and $ \gamma_{\text{FT}}$ are hyperparemters of the Focal Tversky loss $\mathcal{L}_\text{FT}$. The Cross-Entropy loss \eqref{eq:bce} and Dice loss \eqref{eq:dice} are commonly used in semantic segmentation. To address the challenge of underrepresented instance classes, the Focal Tversky loss \eqref{eq:focal_tversky}, a generalization of the Tversky loss, is used. The Focal Tversky loss places greater emphasis on accurately classifying underrepresented instances by assigning higher weights to those samples. This weighting  enhances the model's capacity to handle class imbalance and focuses its learning on the more challenging regions of the segmentation task. 

\subsection{Postprocessing}
\label{subsec:postprocessing}
As the network does not directly provide a semantic instance segmentation with separated nuclei, postprocessing is necessary to obtain accurate results. This involves several steps, including merging the information from the different branches, separating overlapping nuclei to ensure accurate individual segmentation, and determining the nuclei class based on the nuclei type map. Moreover, when performing inference on whole gigapixel WSI, a fusion mechanism is necessary. Due to the significant size of WSIs, inference needs to be performed on image patches extracted from them using a sliding-window approach. The segmentation results obtained from these patches must be assembled to generate a segmentation map of the entire WSI. The postprocessing methods are therefore explained in the following two paragraphs, starting with the segmentation of a single patch followed by its composition into a segmentation output for the entire WSI.
     
\paragraph{Nuclei Separation and Classification} 
To separate adjacent and overlapping nuclei from each other, we utilize HoVer-Net's validated postprocessing pipeline. This involves computing the gradients of the horizontal and vertical distance maps to capture transitions between nuclei boundaries and the boundary between nuclei and the background. At these transition points significant value changes occur in the gradient. The Sobel operator (edge detection filter) is then applied to identify regions with substantial differences in neighboring pixels within the distance maps. Finally, a marker-controlled watershed algorithm is employed to generate the final boundaries. \\
To calculate the nuclei class, the output of the separated nuclei is merged with the nuclei type predictions. For this purpose, majority voting is performed in the nuclei region using the NT prediction map with the majority class assigned to all nuclei pixels \citep{Hovernet}. \\
The STARDIST and CPP-Net decoder methods, on the other hand, use non-maximum suppression (NMS) to prune redundant polygons that likely represent the same object \citep{stardist, stardist2}. We use this approach when testing CellViT with STARDIST and CPP-Net decoders. In difference to STARDIST, the CPP-Net approach uses the refined radial distances as input for the NMS. The nuclei classes are then again assigned to the resulting binary polygons via majority voting. 

\paragraph{Inference}
The encoder ViT offers a significant advantage for performing inference on gigapixel WSI over CNNs based U-Nets. Its capability to process input sequences of arbitrary length, constrained only by memory consumption and positional embedding interpolation, allows for increased input image sizes during inference. It is important to note that positional embedding interpolation must be considered when scaling the input images. 
In preliminary experiments on the MoNuSeg dataset (see Sec. \ref{subsec:generalization_performance}), we found that our network achieves equal performance when inferring on a single $1024 \times 1024~\text{px}$ patch compared to cutting the same patch into $256 \times 256~\text{px}$ sub-patches with an overlap of $64~\text{px}$. Based on these findings, we have chosen to perform WSI inference using $1024 \times 1024~\text{px}$ large patches with a $64~\text{px}$ overlap. Due to the high computational overhead, it is not feasible to keep the segmentation results of the entire WSI in memory. Consequently, we process and merge only the overlapping nuclei during postprocessing. By utilizing just a small overlap in the inference patches relative to the patch size, the postprocessing effort is reduced. To efficiently store the results in a structured and readable format, as well as for compatibility with software such as QuPath \citep{qupath}, the nuclei predictions for an entire WSI are exported in a JSON file. Each nucleus is represented by several parameters, including the nuclei class, bounding-box coordinates, shape polygon of the boundaries, and the center of mass for detection location. In the Appendix, we provide example visualizations of the prediction results from an internal esophageal adenocarcinoma and melanoma cohort, imported into QuPath (see Fig. \ref{fig:qupath_import}). This approach ensures the accessibility of the instance segmentation results for further analysis and visualization. \\
Moreover, for each detected nuclei $\hat{y}$, we store the corresponding embedding token $\bm{z}^{\hat{y}}_{L} \in \mathbb{R}^{D}$. Importantly, as the cell embedding vectors can be directly extracted during the forward pass and are spatially linked to each nuclei $\hat{y}$, there is no need for an additional forward pass on cropped image patches of the detected cells, again saving inference time. If a nucleus is associated with multiple tokens, we average over all token embeddings in which the nucleus is located. The cell embedding can be used as extracted cell-features for downstream DL algorithms addressing problems such as disease prediction, treatment response, and survival prediction. 

\section{Experimental Setup}
\subsection{Datasets}
\label{subsec:datasets}

\begin{figure}
    \centering
    \includegraphics[width=\columnwidth]{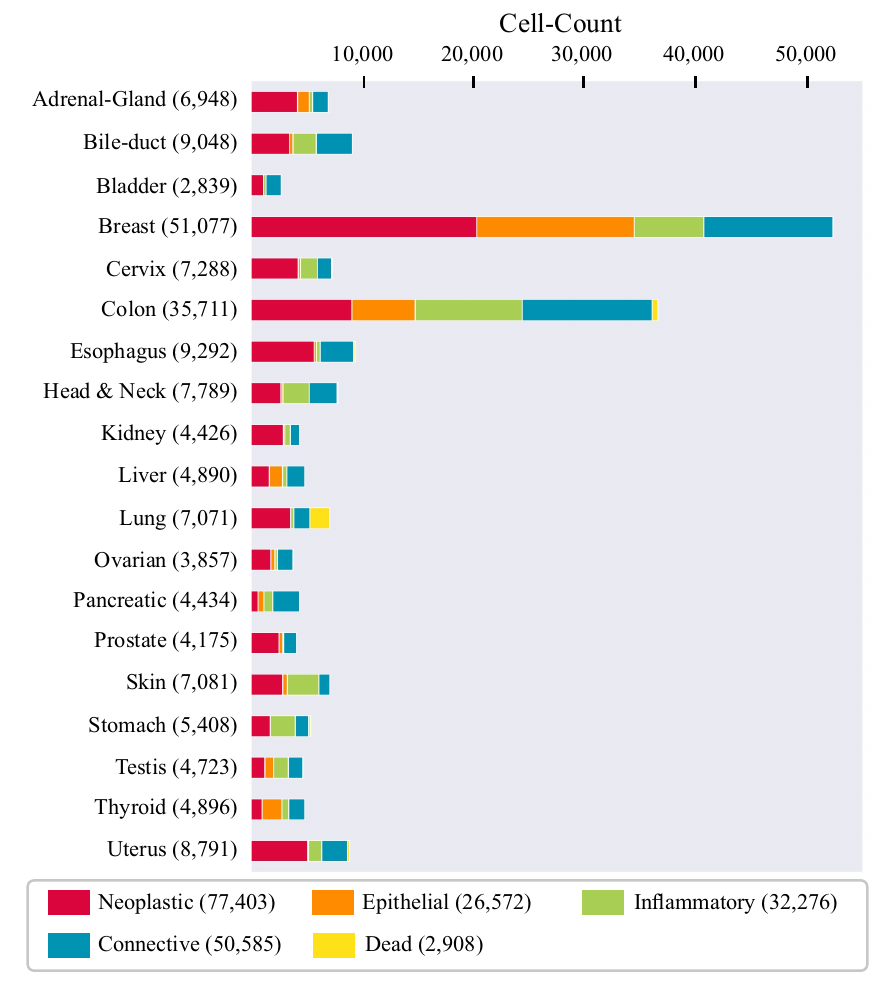}
    \caption{PanNuke nuclei distribution overview for each of the nineteen tissue types, sorted by the total number of nuclei inside the tissue. The total number of nuclei within a tissue type is given in parentheses. Adapted from \citep{pannuke}.}
    \label{fig:pannuke_distribution}
\end{figure}

\paragraph{PanNuke}
We use the PanNuke dataset as the main dataset to train and evaluate our model. The dataset contains $189{,}744$ annotated nuclei in $7{,}904$ $256 \times 256~\text{px}$ images of 19 different tissue types and 5 distinct cell categories, as depicted in Fig. \ref{fig:pannuke_distribution}. Cell-images were captured at a magnification of $\times 40$ with a resolution of $0.25~\si{\micro\meter \per px}$. The dataset is highly imbalanced, especially the nuclei class of dead cells is severely underrepresented, as apparent in the nuclei and tissue class statistics (see Fig. \ref{fig:pannuke_distribution}). PanNuke is regarded as one of the most challenging datasets to perform the simultaneous nuclei instance segmentation task \citep{tsfd}.

\paragraph{MoNuSeg}
The MoNuSeg\citep{monuseg_1, monuseg_2} dataset serves as an additional dataset for nuclei segmentation. In contrast to PanNuke, the dataset is much smaller and does not divide the nuclei into different classes. For this work, we only use the test dataset of MoNuSeg to evaluate our model. The test dataset consists of 14 images with a resolution of $1000 \times 1000~\text{px}$, acquired at $\times 40$ magnification with $0.25~\si{\micro\meter \per px}$. In total, the test dataset contains more than 7000 annotated nuclei across the seven organ types kidney, lung, colon, breast, bladder, prostate, and brain at several disease states (benign and tumors at different stages). Since no nuclei labels are included, the dataset cannot be used for evaluating classification performance.
To process the dataset more effectively with our ViT-based networks with a token size of $16~\text{px}$, we resized the data to a size of $1024 \times 1024~\text{px}$. Due to the sufficient patch size of the original data, we also created a $\times 20$ dataset with $0.50~\si{\micro\meter \per px}$ resolution, where the patch size is $512 \times 512~\text{px}$ accordingly.

\paragraph{CoNSeP}
We utilized the colorectal nuclear segmentation and phenotypes (CoNSeP) dataset by \citet{Hovernet} to analyze extracted cell embeddings (see Sec. \ref{subsec:postprocessing}) of detected cells on an external validation dataset. This dataset comprises 41 H\&E-stained colorectal adenocarcinoma WSI at a resolution of $0.25~\si{\micro\meter \per px}$ and an image size of $1000 \times 1000~\text{px}$, which we rescale to $1024 \times 1024~\text{px}$ similar to the MoNuSeg data. The dataset exhibits significant diversity, encompassing stromal, glandular, muscular, collagen, adipose, and tumorous regions, along with various types of nuclei derived from originating cells: normal epithelial, dysplastic epithelial, inflammatory, necrotic, muscular, fibroblast, and miscellaneous nuclei, including necrotic and mitotic cells.

\subsection{Experiments}
\label{subsec:baseline_experiments}
In this study, we conducted two experiments on the PanNuke dataset and one on the MoNuSeg dataset to assess algorithms performance. We additionally used an internal dataset for comparing inference speed time. Given the higher clinical relevance of the detection task over achieving the optimal segmentation quality, we \textbf{(1)} performed an ablation study on PanNuke to determine the most suitable network architecture for nuclei detection. We compared the performance of pre-trained models (see Sec. \ref{subsec:model_training}) against randomly initialized models and explored the impact of regularization techniques such as data augmentation, loss functions, and customized oversampling, as well as comparing the HoVer-Net decoder method to the STARDIST and CPP-Net decoder methods in our UNETR-structure. Based on these investigations, we identified the best models, which were \textbf{(2)} subsequently evaluated for segmentation quality. To assess both detection and segmentation performance, we compared our models with multiple baseline architectures, namely DIST \citep{dist}, Mask-RCNN \citep{mask_rcnn}, Micro-Net \citep{micro_net}, HoVer-Net \citep{Hovernet}, TSFD-Net \citep{tsfd}, and CPP-Net \citep{cpp_net}. We also re-trained the STARDIST model with a ResNet50 \citep{resnet50} backbone and the hyperparameters of \citet{cpp_net} to retrieve comparable detection results.
For comparison, we conducted our experiments using the same three-fold cross-validation (CV) splits provided by the PanNuke dataset organizers and report the averaged results over all three splits. It is worth mentioning that all the comparison models we evaluate in this study adhere to the same evaluation scheme for the PanNuke dataset, with one exception. The TSFD-Net publication reports results based on an 80-20 train-test split, making their results more optimistic. Nevertheless, we include their results for the purpose of comparison. \\
As a third experiment \textbf{(3)}, we evaluated our models trained on PanNuke on the publicly available 14 test images of the MoNuSeg dataset to test generalizability. The dataset serves a second purpose next to generalization: We compare various input image sizes and assess the performance of our inference pipeline outlined in Section \ref{subsec:postprocessing}. In this context, we evaluate the performance using two scenarios - one involving an uncropped MoNuSeg slide with $1024~\text{px}$ input patch size and the other using cropped $256~\text{px}$ input images. Additionally, we investigate the impact of our overlapping strategy with a 64-pixel overlap, focusing on the $256~\text{px}$ input size. \\
To analyze the cell embeddings $\bm{z}^{\hat{y}}_{L}$ for detected nuclei with our CellViT models, we utilize the CoNSeP dataset \textbf{(4)}. To achieve this, we perform inference with the pre-trained PanNuke models on the CoNSeP images ($1024~\text{px}$ input patch size) and extract the token embeddings $\bm{z}^{\hat{y}}_{L}$ for each nuclei $\hat{y}$ from the last Transformer block that are spatially associated with $\hat{y}$. Subsequently, we employ the Uniform Manifold Approximation and Projection (UMAP) method for dimension reduction to transform the cell embedding vectors (of the 27 training images) into a two-dimensional representation, which can be visualized in a two-dimensional scatter plot. We additionally trained a linear classifier on top of the cell embeddings (extracted from the 27 training images) to classify the detected cells into the CoNSeP nuclei classes and tested the classifier on the cell embeddings of the cells from the 14 test images. \\
Finally, to compare the inference runtime \textbf{(5)}, we collected a diverse dataset of 10 esophageal WSIs with tissue areas ranging from $2.79~\si{\milli\meter}^2$ to $74.07~\si{\milli\meter}^2$. We measured the inference runtime for the HoVer-Net model, as well as for the $\text{CellViT}_{256}$ and CellViT-SAM-H models with $256~\text{px}$ and $1024~\text{px}$ patch input size and an overlap of $64~\text{px}$. For each WSI, we repeated the process three times and averaged the runtime results. 

\subsection{Evaluation Metrics}
\label{subsec:evaluation_metrics}

\paragraph{Nuclear Instance Segmentation Evaluation}
Usually, the Dice coefficient (DICE) or the Jaccard index are used as evaluation metrics for semantic segmentation. However, as \citet{Hovernet} have already shown, these two metrics are insufficient for evaluating nuclear instance segmentation as they did not account for the detection quality of the nuclei. Therefore, a metric is needed that assess the following three requirements (see \citet{Hovernet}):
\begin{enumerate}
    \item Separate the nuclei from the background \label{metric:task1}
    \item Detect individual nuclei instances and separate overlapping nuclei
    \item Segment each instance 
\end{enumerate}
These three requirements cannot be evaluated with the Jaccard index and the DICE score, as they just satisfy requirement (\ref{metric:task1}). In line with \citep{Hovernet} and the PanNuke dataset evaluation recommendations \citep{pannuke}, we use the panoptic quality ($PQ$) \citep{kirillov_pq} to quantify the instance segmentation performance. The $PQ$ us defined as
\begin{align}
    PQ=\underbrace{\frac{\left | TP \right |}{\left | TP \right | + \frac{1}{2}\left | FP \right | + \frac{1}{2}\left | FN \right |}}_{\text{Detection Quality (DQ)}} \times \underbrace{\frac{\sum_{(y,\hat{y})\in{TP}} IoU(y,\hat{y})}{\left |  TP \right |}}_{\text{Segmentation Quality (SQ)}}, 
\end{align}
with $IoU(y,\hat{y})$ denoting the intersection-over-union \citep{kirillov_pq}.
In this equation, $y$ denotes a ground-truth (GT) segment, and $\hat{y}$ denotes a predicted segment, 
with the pair $(y,\hat{y})$ being a unique matching set of one ground-truth segment and one predicted segment. As \citet{kirillov_pq} proved, each pair of segments $(y,\hat{y})$, i.e., each pair of true and predicted nuclei, in an image is unique if $IoU(y,\hat{y}) > 0.5$ is satisfied. For each class, the unique matching of $(y,\hat{y})$ splits the predicted and the GT segments into three sets:
\begin{itemize}
    \item True Positives (TP): Matched pairs of segments, \\ i.e., correctly detected instances
    \item False Positives (FP): Unmatched predicted segments, \\ i.e., predicted instances without matching GT instance
    \item False negatives (FN): Unmatched GT segments, \\ i.e., GT instances without matching predicted instance
\end{itemize}
The $PQ$ score can be intuitively decomposed into two parts, the detection quality similar to the $F_1$ score commonly used in classification and detection scenarios, and the segmentation quality as the average $IoU$ of matched segments \citep{Hovernet, kirillov_pq}. To ensure a fair comparison, we use binary $PQ$ ($bPQ$) pretending that all nuclei belong to one class (nuclei vs. background) and the more challenging multi-class PQ ($mPQ$), taking the nuclei class into account. In doing so for $mPQ$, we calculate the $PQ$ independently for each nuclei class and subsequently average the results over all classes \citep{pannuke}. 

\paragraph{Nuclear Classification Evaluation}
To evaluate the detection quality of our model, we employ commonly used detection metrics. Similar to the approach used in the $PQ$-score for nuclear instance segmentation evaluation, we split GT and predicted instances into TPs, FPs, and FNs.  We use the conventional detection metrics precision ($P_\text{d}$), recall ($P_\text{d}$) and the ($F_{1,\text{d}}$)-score as a harmonic mean between precision and recall. The index '\textit{d}' indicates that these are the scores for the entire binary nuclei detection over all classes $c$. Thus, the binary detection scores are defined as follows:
\begin{align*}
    F_{1,\text{d}} & = \frac{2TP_\text{d}}{2TP_\text{d} + FP_\text{d} + FN_\text{d}} \\
    P_\text{d}& = \frac{TP_\text{d}}{TP_\text{d}+FP_\text{d}} \\
    R_\text{d} & = \frac{TP_\text{d}}{TP_\text{d}+FN_\text{d}}
\end{align*}
We further break down $TP_d$ into correctly classified instances of class $c$ ($TP_c$), false positives of class $c$ ($FP_c$) and false negatives of class $c$ ($FN_c$) to derive cell-type specific scores. We then define the $F_{1,c}$-score, precision ($P_c$) and recall ($R_c$) of each nuclei class $c$ as 
\begin{align*}
    F_{1,c} & = \frac{2(TP_c+TN_c)}{2(TP_c+TN_c) + 2FP_c + 2FN_c + FP_\text{d} + FN_\text{d}}, \\
    P_c & = \frac{TP_c+TN_c}{TP_c+TN_c + 2FP_c + FP_\text{d}}, \\
    R_c & = \frac{TP_c+TN_c}{TP_c+TN_c + 2FN_c + FN_\text{d}}. 
\end{align*}
In order to prioritize the classification of different nuclear types, we incorporated an additional weighting factor for the nuclei classes, as suggested in the official PanNuke evaluation metrics \citep{pannuke,Hovernet},  Since we cannot use the $IoU(y,\hat{y})>0.5$ criterion to find matching instances $(y,\hat{y})$ between GT-instances and predictions for the detection task, we use the methodology of \citet{siri_2016} and define a match $(y,\hat{y})$ if both centers of mass are within a radius of $6~\text{px}$ ($0.50~\si{\micro\meter \per px}$) and $12~\text{px}$ ($0.25~\si{\micro\meter \per px}$), respectively.
\begin{table*}[!th]
    \centering
    \caption{Precision ($P$), Recall ($R$) and $F_1$-score ($F_1$) for detection and classification across the three PanNuke splits for each nuclei type. The centroid of each nucleus was used for computing detection metrics for segmentation networks.
    *TSFD-Net was not evaluated on the official three-fold splits of the PanNuke dataset and left out by the comparison **Model re-trained by ourselves ***Models trained on downscaled $0.50~\si{\micro\meter \per px}$ PanNuke images}
    \label{tab:detection}
    \resizebox{\textwidth}{!}{%
    \begin{tabular}{llllllllllllllllllllllllll}
    \toprule
                                               {Model} & {Decoder} & {Hyperparameters} & \multicolumn{3}{l}{Detection}                 &  & \multicolumn{19}{l}{Classification}                                                                                                                                                                                                                       \\ \cmidrule(l){8-26} 
                        &  & &             &               &               &  & \multicolumn{3}{l}{Neoplastic}                &  & \multicolumn{3}{l}{Epithelial}                &  & \multicolumn{3}{l}{Inflammatory}              &  & \multicolumn{3}{l}{Connective}                &  & \multicolumn{3}{l}{Dead}                      \\ \cmidrule(lr){8-10} \cmidrule(lr){12-14} \cmidrule(lr){16-18} \cmidrule(lr){20-22} \cmidrule(l){24-26} 
                                                                                    &           &         & $P_{\text{d}}$& $R_{\text{d}}$& $F_{1,\text{d}}$&  & $P_{\text{Neo}}$   & $R_{\text{Neo}}$ & $F_{\text{1,Neo}}$ &  & $P_{\text{Epi}}$   & $R_{\text{Epi}}$  & $F_{1,\text{Epi}}$    &  & $P_{\text{Inf}}$   & $R_{\text{Inf}}$  & $F_{1,\text{Inf}}$    &  & $P_{\text{Con}}$   & $R_{\text{Con}}$  & $F_{1,\text{Con}}$    &  & $P_{\text{Dead}}$  & $R_{\text{Dead}}$ & $F_{1,\text{Dead}}$ \\ \midrule
    DIST                                                                            &           &         & 0.74          & 0.71          & 0.73            &  & 0.49               & 0.55             & 0.50               &  & 0.38               & 0.33              & 0.35                  &  & 0.42               & 0.45              & 0.42                  &  & 0.42               & 0.37              & 0.39                  &  & 0.00               & 0.00              & 0.00                \\
    Mask-RCNN                                                                       &           &         & 0.76          & 0.68          & 0.72            &  & 0.55               & 0.63             & 0.59               &  & 0.52               & 0.52              & 0.52                  &  & 0.46               & 0.54              & 0.50                  &  & 0.42               & 0.43              & 0.42                  &  & 0.17               & 0.30              & 0.22          \\
    Micro-Net                                                                       &           &         & 0.78          & 0.82          & 0.80            &  & 0.59               & 0.66             & 0.62               &  & 0.63               & 0.54              & 0.58                  &  & 0.59               & 0.46              & 0.52                  &  & 0.50               & 0.45              & 0.47                  &  & 0.23               & 0.17              & 0.19          \\
    HoVer-Net                                                                       &           &         & 0.82          & 0.79          & 0.80            &  & 0.58               & 0.67             & 0.62               &  & 0.54               & 0.60              & 0.56                  &  & 0.56               & 0.51              & 0.54                  &  & 0.52               & 0.47              & 0.49                  &  & 0.28               & 0.35              & 0.31          \\
    TSFD-Net*                                                                       &           &         & 0.84          & 0.87          & 0.85            &  & 0.60               & 0.71             & 0.65               &  & 0.56               & 0.58              & 0.57                  &  & 0.59               & 0.58              & 0.57                  &  & 0.55               & 0.49              & 0.53                  &  & 0.33               & 0.40              & 0.43          \\
    STARDIST (ResNet50) **                                                          & STARDIST  & CPP-Net & 0.85          & 0.80          & 0.82            &  & 0.69               & 0.69             & 0.69               &  & 0.73               & 0.68              & 0.70                  &  & 0.62               & 0.53              & 0.57                  &  & 0.54               & 0.49              & 0.51                  &  & 0.39               & 0.09              & 0.10          \\
    STARDIST (ResNet50) **                                                          & STARDIST  & CellViT     & 0.85          & 0.79          & 0.82            &  & 0.70               & 0.66             & 0.68               &  & 0.71               & 0.66              & 0.68                  &  & 0.58               & 0.58              & 0.58                  &  & 0.54               & 0.49              & 0.51                  &  & 0.39               & 0.34              & 0.36          \\ \midrule

    $\text{CellViT}_{256}-\text{Raw}$                                               & HoVer-Net & CellViT     & 0.80          & 0.77          & 0.78            &  & 0.61               & 0.64             & 0.63               &  & 0.63               & 0.59  	        & 0.61                  &  & 0.55               & 0.46              & 0.50                  &  & 0.45               & 0.43              & 0.44                  &  & 0.43               & 0.16              & 0.23          \\
    $\text{CellViT}_{256}-\text{Over}$                                              & HoVer-Net & CellViT     & 0.79          & 0.78          & 0.78            &  & 0.62               & 0.63             & 0.62               &  & 0.65               & 0.59  	        & 0.62                  &  & 0.54               & 0.47              & 0.50                  &  & 0.44               & 0.45              & 0.44                  &  & \textbf{0.46}      & 0.16              & 0.24          \\
    $\text{CellViT}_{256}-\text{Aug}$                                               & HoVer-Net & CellViT     & 0.83          & 0.82          & 0.82            &  & 0.70               & 0.69             & 0.69               &  & 0.68               & 0.71  	        & 0.69                  &  & 0.58               & \textbf{0.59}     & 0.58                  &  & 0.54               & 0.51              & 0.52                  &  & 0.38               & 0.35              & 0.36          \\
    $\text{CellViT}_{256}-\text{No-FC}$                                             & HoVer-Net & CellViT     & 0.82          & \textbf{0.83} & 0.82            &  & 0.69               & \textbf{0.70}    & 0.69               &  & 0.70               & 0.69  	        & 0.70                  &  & 0.58               & 0.58              & 0.58                  &  & 0.53               & 0.51              & 0.52                  &  & 0.40               & 0.33              & 0.36          \\ 
    CellViT-Random (no pre-train)                                                   & HoVer-Net & CellViT     & 0.79          & 0.81          & 0.80            &  & 0.63               & 0.65             & 0.64               &  & 0.63               & 0.62  	        & 0.72                  &  & 0.54               & 0.57              & 0.55                  &  & 0.49               & 0.46              & 0.48                  &  & 0.30               & 0.34              & 0.31          \\ \hline \rowcolor[rgb]{0.753,0.753,0.753}  & & & & & & & & & & & & & & & & & & & & & & & & & \\ [-2ex]
    
    \rowcolor[rgb]{0.753,0.753,0.753} $\text{CellViT}_{256}$                        & HoVer-Net & CellViT     & 0.83          & 0.82          & 0.82            &  & 0.69               & \textbf{0.70}    & 0.69               &  & 0.68               & 0.71  	        & 0.70                  &  & 0.59               & 0.58              & 0.58                  &  & 0.53               & 0.51              & 0.52                  &  & 0.39               & 0.35              & 0.37          \\
    \rowcolor[rgb]{0.753,0.753,0.753} CellViT-SAM-B                                 & HoVer-Net & CellViT     & 0.83          & 0.82          & \textbf{0.83}   &  & 0.70               & \textbf{0.70}    & 0.70               &  & 0.70               & 0.72  	        & 0.71                  &  & 0.59               & 0.58              & \textbf{0.59}         &  & 0.54               & \textbf{0.52}     & \textbf{0.53}         &  & \textbf{0.46}      & 0.29              & 0.36          \\ 
    \rowcolor[rgb]{0.753,0.753,0.753} CellViT-SAM-L                                 & HoVer-Net & CellViT     & 0.84          & 0.82          & \textbf{0.83}   &  & 0.71               & \textbf{0.70}    & 0.70               &  & 0.71               & 0.72  	        & 0.72                  &  & 0.59               & 0.58              & 0.58                  &  & 0.54               & \textbf{0.52}     & \textbf{0.53}         &  & 0.42               & 0.36              & \textbf{0.39} \\ 
    \rowcolor[rgb]{0.753,0.753,0.753} CellViT-SAM-H                                 & HoVer-Net & CellViT     & 0.84          & 0.81          & \textbf{0.83}   &  & 0.72               & 0.69             & \textbf{0.71}      &  & 0.72               & \textbf{0.73}     & \textbf{0.73}         &  & 0.59               & 0.57              & 0.58                  &  & 0.55               & \textbf{0.52}     & \textbf{0.53}         &  & 0.43               & 0.32              & 0.36          \\ \hline & & & & & & & & & & & & & & & & & & & & & & & & & \\ [-2ex]

    $\text{CellViT}_{256}$                                                          & STARDIST  & CPP-Net & 0.84          & 0.75          & 0.79            &  & 0.64               & 0.60             & 0.62               &  & 0.65               & 0.56  	        & 0.60                  &  & \textbf{0.64}      & 0.45              & 0.52                  &  & \textbf{0.58}      & 0.47              & 0.47                  &  & 0.30               & 0.27              & 0.28          \\ 
    $\text{CellViT}_{256}$                                                          & STARDIST  & CellViT     & 0.83          & 0.79          & 0.81            &  & 0.71               & 0.65             & 0.68               &  & 0.68               & 0.68  	        & 0.68                  &  & 0.59               & 0.57              & 0.58                  &  & 0.52               & 0.49              & 0.50                  &  & 0.37               & \textbf{0.38}     & 0.37          \\ 
    CellViT-SAM-H                                                                   & STARDIST  & CPP-Net & 0.84          & 0.78          & 0.81            &  & 0.68               & 0.66             & 0.67               &  & 0.71               & 0.62  	        & 0.66                  &  & 0.57               & 0.57              & 0.57                  &  & 0.54               & 0.45              & 0.49                  &  & 0.36               & 0.32              & 0.32          \\ 
    CellViT-SAM-H                                                                   & STARDIST  & CellViT     & 0.84          & 0.80          & 0.82            &  & 0.72               & 0.68             & 0.70               &  & \textbf{0.74}      & 0.71  	        & 0.72                  &  & 0.60               & 0.57              & 0.58                  &  & 0.53               & 0.51              & 0.52                  &  & 0.44               & 0.34              & 0.38          \\ \midrule

    $\text{CellViT}_{256}$                                                          & CPP-Net   & CPP-Net & 0.85          & 0.76          & 0.80            &  & 0.69               & 0.62             & 0.65               &  & 0.70               & 0.62  	        & 0.65                  &  & 0.57               & 0.55              & 0.56                  &  & 0.53               & 0.46              & 0.49                  &  & 0.32               & \textbf{0.38}     & 0.33          \\ 
    $\text{CellViT}_{256}$                                                          & CPP-Net   & CellViT     & \textbf{0.87} & 0.76          & 0.81            &  & 0.73               & 0.64             & 0.68               &  & 0.71               & 0.65  	        & 0.68                  &  & 0.58               & 0.57              & 0.58                  &  & 0.55               & 0.47              & 0.51                  &  & 0.37               & 0.37              & 0.37          \\ 
    CellViT-SAM-H                                                                   & CPP-Net   & CPP-Net & 0.86          & 0.78          & 0.82            &  & 0.72               & 0.67             & 0.70               &  & 0.73               & 0.68  	        & 0.70                  &  & 0.62               & 0.55              & 0.58                  &  & 0.55               & 0.50              & 0.52                  &  & 0.27               & 0.14              & 0.18          \\ 
    CellViT-SAM-H                                                                   & CPP-Net   & CellViT     & \textbf{0.87} & 0.78          & 0.82            &  & \textbf{0.74}      & 0.67             & 0.70               &  & \textbf{0.74}      & 0.70  	        & 0.72                  &  & 0.60               & 0.57              & 0.58                  &  & 0.57               & 0.49              & \textbf{0.53}         &  & 0.41               & 0.36              & 0.38          \\ \midrule

    $\text{CellViT}_{256}$ ($0.50~\si{\micro\meter \per px}$)***                    & HoVer-Net & CellViT     & 0.86          & 0.60          & 0.71            &  & 0.72               & 0.59             & 0.65               &  & 0.71               & 0.58  	        & 0.64                  &  & 0.60               & 0.38              & 0.47                  &  & 0.53               & 0.32              & 0.40                  &  & 0.43               & 0.04              & 0.07          \\ 
    CellViT-SAM-H ($0.50~\si{\micro\meter \per px}$)***                             & HoVer-Net & CellViT     & 0.88          & 0.63          & 0.73            &  & 0.74               & 0.62             & 0.67               &  & 0.74               & 0.61  	        & 0.67                  &  & 0.60               & 0.42              & 0.49                  &  & 0.56               & 0.34              & 0.42                  &  & 0.49               & 0.04              & 0.08          \\ \bottomrule
    \end{tabular}%
    }
\end{table*}
\vspace{4mm}
\subsection{Model Training}
\label{subsec:model_training}

\paragraph{Oversampling}
Even though the PanNuke dataset has around $200{,}000$ annotated nuclei, they are distributed just across a limited number of $8{,}000$ patches with $256\times 256~\text{px}$ patch size. Furthermore, there is a substantial class imbalance among tissue types and nucleic classes
(see Fig. \ref{fig:pannuke_distribution}). Thus, we developed a new oversampling strategy based on class weightings to balance both tissue classes and nuclei classes. For each patch $i$ in the training dataset with $N_\text{Train}$ training samples, we calculate the sampling weights for the tissue class and the cell class with 
\begin{align}
    \label{eq:sampling_1}
    p_{i}(\gamma_\text{s}) = \frac{w_\text{Tissue}(i, \gamma_\text{s})}{\displaystyle\max_{j \in \left[1, N_\text{Train}\right]} w_\text{Tissue}(j, \gamma_\text{s})} 
    + 
    \frac{w_\text{Cell}(i, \gamma_\text{s})}{\displaystyle\max_{\substack{j \in \left[1, N_\text{Train}\right]}} w_\text{Cell}(j, \gamma_\text{s})},
\end{align} where $w_\text{Tissue}(i, \gamma_\text{s})$ is a weight factor for the tissue class and $w_\text{Cell}(i, \gamma_\text{s})$ for the nuclei class.
The parameter $\gamma_\text{s} \in [0, 1]$ is a weighting factor that determines the strength of the oversampling. A $\gamma_\text{s}$ value of 0 indicates no oversampling, while  $\gamma_\text{s}=1$ corresponds to maximum balancing. To ensure neither $w_\text{Tissue}(i, \gamma_\text{s})$ nor  $w_\text{Cell}(i, \gamma_\text{s})$ dominates the sampling, normalization is applied to both summands in eq. \eqref{eq:sampling_1}.
The calculation of the weighting factor of the tissue class can be calculated directly via
\begin{align}
    \label{eq:tissue_scaling}
    w_\text{Tissue}(i, \gamma_\text{s}) = \frac{N_\text{Train}}{\gamma_\text{s}\left(\displaystyle{\sum_{j \in \left[1, N_\text{Train}\right] | c_{\text{T},j} = c_{\text{T},i}}} 1 \right ) +(1-\gamma_\text{s})N_\text{Train}}
\end{align}
as each patch can only belong to one tissue class denoted by $c_{\text{T},i}$.\\
For cell weighting, it must be considered that each patch can contain multiple nuclei from different cell classes. Therefore, we create a binary vector $\bm{c_i}\in \left\{0,1\right\}^{C}$, where each entry is set to 1 for each existing nuclei type $c$ in the patch. To get a reference value for scaling similar to eq. \eqref{eq:tissue_scaling}, we calculate $N_{\text{Cell}} = \sum_{i=1}^{N_\text{Train}} \left \| \bm{c}_i \right \|_1$. The cell weighting for each training image $i$ is then calculated by
\begin{align*}
    w_\text{Cell}(i, \gamma_\text{s}) = (1-\gamma_\text{s})+\gamma_\text{s}
    \sum_{j=1}^{C}
    c_{ij} \frac{N_{\text{Cell}}}{\gamma_\text{s}\sum_{k=1}^{N_\text{Train}} c_{kj} + (1-\gamma_\text{s})N_{\text{Cell}}},
\end{align*}
with $c_{ij}$ the vector entry of $\bm{c}_i$ at position $j$. The training images are randomly sampled in a training epoch with replacement based on their sampling weights $p_{i}(\gamma_\text{s})$.

\paragraph{Data Augmentation}
In addition to our customized oversampling strategy, we extensively employ data augmentation techniques to enhance data variety and discourage overfitting. We use a combination of the following geometrical and noisy/intensity-based augmentation methods: random 90-degree rotation, horizontal flipping, vertical flipping, downscaling, blurring, gaussian noise, color jittering, superpixel representation of image sections (SLIC), zoom blur, random cropping with resizing and elastic transformations. These augmentation techniques were selected to introduce variations in the shape, orientation, texture, and appearance of the nuclei, enhancing the robustness and generalization capabilities of the model. For detailed information on the augmentation methods utilized, including the selected probabilities and corresponding hyperparameters, please refer to the Appendix.

\paragraph{Optimization and Training Strategy}
We train all our models for 130 epochs and incorporate exponential learning rate scheduling with a scheduling factor of 0.85 to gradually reduce the learning rate during training (denoted as CellViT hyperparameters). To balance our training, we use our modified oversampling strategy with $\gamma_\text{s}=0.85$. For the STARDIST and CPP-Net models, we also conducted experiments using the proposed CPP-Net hyperparameters by \citet{cpp_net}. A complete overview of all hyperparameters, including optimizer, data augmentation, and weighting factors of the loss functions in eq. \eqref{eq:branch_losses} is provided in the Appendix. 
As for the encoder models, we leverage the $\text{ViT}_{256}$-model (ViT-S, $D=384$, $L=12$), which has been pre-trained on histological data (see Sec. \ref{subsec:vision_transformer}). Additionally, we compare the performance with the three pre-trained SAM checkpoints: SAM-B (ViT-B, $D=768$, $L=12$), SAM-L (ViT-L, $D=1024$, $L=24$) and SAM-H (ViT-H, $D=1280$, $L=32$). These checkpoints provide different model sizes and complexities, allowing us to evaluate their respective performance and choose the most suitable one for our task. During training, we initially freeze the encoder weights for the first 25 epochs. After this initial warm-up phase to train the decoder, we proceed to train the entire model including the image encoder. 

\paragraph{Implementation}
All models are implemented in PyTorch 1.13.1. To augment images and masks, we used the Albumentations library \citep{albumentations}.  Other used libraries include the official STARDIST \citep{stardist2}, CPP-Net \citep{cpp_net} and CellSeg-models implementations \citep{cellseg_models_pt}. For the pre-trained $\text{ViT}_{256}$-model, we utilized the ViT-S checkpoint\footnote[1]{\tiny\url{https://github.com/mahmoodlab/HIPT}} provided by \citet{vit256}. As for the SAM-B, SAM-L, and SAM-H models, we use the encoder backbones of each final training stage of SAM \citep{SAM}, published on GitHub\footnote[2]{\tiny\url{https://github.com/facebookresearch/segment-anything}}. All experiments were conducted on an $80~\text{GB}$ NVIDIA A100 GPU with automatic mixed precision. However, it is worth noting that a $48~\text{GB}$ NVIDIA RTX A6000 is also sufficient for the $\text{ViT}_{256}$ and SAM-B model training.

\section{Results}
\label{sec:results}

In the section below, the results for the experiments (1) nuclei detection quality and (2) segmentation quality on PanNuke, (3) generalization performance on the independent MoNuSeg cohort, (4) cell-embedding analysis and (5) inference speed comparisons are given. If not stated otherwise, all models were trained on the PanNuke dataset with a resolution of $0.25~\si{\micro\meter \per px}$.

\vspace{-2mm}
\subsection{Detection Quality on PanNuke}
\label{subsec:ablation_study}

Considering the clinical importance of nuclei detection and classification over achieving the best possible segmentation quality, our ablation study aimed to determine the best model based on the detection results using the PanNuke dataset. Tab. \ref{tab:detection} presents the precision, recall, and $\text{F}_1$-Score for both detection and classification performance across all nuclei classes, including the binary case.
To determine the optimal settings, we evaluated different variations of our network. These include a randomly initialized network (CellViT-Random), networks with pre-trained weights from the $\text{ViT}_{256}$ network (CellViT$_{256}$), and networks with different pre-trained SAM backbones (CellViT-SAM-B, CellViT-SAM-L, CellViT-SAM-H). To ensure comparability, the CellViT-Random network shares the same architecture (ViT-S, $D=384$, $L=12$) as the CellViT$_{256}$ network. All mentioned CellViT model variants were trained using data augmentation and our customized sampling strategy as regularization methods. The decoder network strategies (HoVer-Net, STARDIST or CPP-Net decoder) and hyperparameter settings are given behind the network name in Tab. \ref{tab:detection}. \\
We first analyze the CellViT models with HoVer-Net decoder. Compared to the baseline models, the randomly initialized CellViT-Random network achieves detection results comparable to the HoVer-Net CNN network. However, when using pre-trained encoder networks, we observe a significant performance increase, reaching state-of-the-art performance. We notice a strong increase in $F_1$-scores compared to all existing solutions, especially for the epithelial nuclei class. Both the $\text{ViT}_{256}$ and the three different SAM encoders exhibit significantly better performance, all at a similar level, with the CellViT-SAM-H model as the best solution. Notably, we even outperform purely detection-based methods like Mask-RCNN and all state-of-the-art approaches by a large margin with up to a $26~\%$ increase in the $F_{1,\text{Epi}}$-score of epithelial nuclei. \\
To demonstrate the effect of extensive data augmentation, customized sampling strategy and the Focal Tversky loss, we additionally report the results for a $\text{CellViT}_{256}$ model without regularization ($\text{CellViT}_{256}$-Raw), with oversampling only ($\text{CellViT}_{256}$-Over), with data-augmentation only ($\text{CellViT}_{256}$-Aug) and a model trained with oversampling and all augmentations, but without Focal Tversky loss ($\text{CellViT}_{256}$-No-FC) in Tab. \ref{tab:detection}. Our experiments reveal that data augmentation, in particular, is a crucial regularization method that significantly enhances performance. Specifically, the addition of data augmentation results in a 0.13 increase in the $F_{1,\text{Dead}}$ score for the dead nuclei class compared the the ($\text{CellViT}_{256}$-Raw) model. Oversampling and Focal Tversky loss just lead to minimal improvements in the detection scores. \\
We also tested the STARDIST and CPP-Net decoder structures with the $\text{CellViT}_{256}$ and CellViT-SAM-H model with our hyperparameters and the CPP-Net hyperparameters suggested by \citet{cpp_net}. These models usually achieve higher precision values but often a significantly lower recall and a lower F1 score than the models with the HoVer-Net decoder architecture. As an extension of the STARDIST method, the CPP-Net decoder achieves slightly better results. Overall, the models achieve better detection results than comparable CNN-based SOTA networks and outperform the ResNet50-based STARDIST model, but are inferior to our suggested models with HoVer-Net decoder architecture. The results also reveal that our hyperparameters provide better detection performance.  \\
In addition to the provided dataset resolution of $0.25~\si{\micro\meter \per px}$, we performed training and evaluation for the two best model variants $\text{CellViT}_{256}$ and CellViT-SAM-H on downscaled PanNuke data (from $256 \times 256$ to $128 \times 128~\text{px}$ patch size), resulting in $0.50~\si{\micro\meter \per px}$ resolution. The results are presented in the last two rows of Tab. \ref{tab:detection}. The downsizing leads to a substantial drop in performance compared to the $0.25~\si{\micro\meter \per px}$ networks, with detection results approaching the baseline models. Notably, the recall of individual classes significantly decreases (by an average of $-0.20$). In particular, the recall for the dead nuclei class drops to 0.04, indicating that this class is almost not detected at all. Interestingly, the precision increases minimally or remains almost the same compared to our best $0.25~\si{\micro\meter \per px}$ models. We conclude that despite detecting significantly fewer nuclei, when a nucleus is identified and classified correctly, it corresponds to the true nucleus class with high accuracy for most classes. \\
For subsequent investigations, we decided to further just consider the $\text{CellViT}_{256}$ and CellViT-SAM-H models to enable a comparison between in-domain and out-of-domain pre-training. 

\begin{table}[!t]
      \centering
      \caption{Average $PQ$ across the three PanNuke splits for each nuclear category on the PanNuke dataset. *TSFD-Net was not evaluated on the official three-fold splits of the PanNuke dataset and left out by the comparison. **Model re-trained by ourselves  ***Models trained on downscaled $0.50~\si{\micro\meter \per px}$ PanNuke images. \\ Abbreviations: Decoder (Dec.), Hyperparameter (HP.), HoVer-Net (HV), STARDIST (SD), CPP-Net (CPP).
      }
      \label{tab:pq_nuclei}
      \resizebox{\columnwidth}{!}{%
          \begin{tabular}{l@{\hspace{2pt}}l@{\hspace{2pt}}lc@{\hspace{2pt}}c@{\hspace{3pt}}c@{\hspace{3pt}}c@{\hspace{3pt}}c}
          \toprule
                                                                                                    {Model}   & {Dec.}  & {HP.}          & \small Neoplastic    & \small Epithelial  & \small Inflammatory   & \small Connective     & \small Dead       \\ \midrule
          \small{DIST}                                                                                 &       &                  & 0.439                & 0.290              & 0.343                 & 0.275                 & 0.000             \\
          Mask-RCNN                                                                                    &       &                  & 0.472                & 0.403              & 0.290                 & 0.300                 & 0.069             \\
          Micro-Net                                                                                    &       &                  & 0.504                & 0.442              & 0.333                 & 0.334                 & 0.051             \\
          HoVer-Net                                                                                    &       &                  & 0.551                & 0.491              & 0.417                 & 0.388                 & 0.139             \\
          TSFD-Net*                                                                                    &       &                  & 0.572                & 0.566              & 0.453                 & 0.423                 & 0.214             \\  
          STARDIST (RN50)**                                                                            & SD    & CPP              & 0.564                & 0.543              & 0.398                 & 0.388                 & 0.024             \\ 
          STARDIST (RN50)**                                                                            & SD    & CellViT              & 0.547                & 0.532              & \textbf{0.424}        & 0.380                 & 0.123             \\ \hline   \rowcolor[rgb]{0.753,0.753,0.753} & & & & & & & \\ [-2ex]
          \rowcolor[rgb]{0.753,0.753,0.753} CellViT-SAM-H                                                                                & HV    & CellViT              & \textbf{0.581}       & \textbf{0.583}     & 0.417                 & \textbf{0.423}        & 0.149             \\
          \rowcolor[rgb]{0.753,0.753,0.753} $\text{CellViT}_{256}$                                                                       & HV    & CellViT              & 0.567                & 0.559              & 0.405                 & 0.405                 & 0.144             \\ \hline & \\ [-2ex]

          $\text{CellViT}_{256}-\text{Raw}$                                                            & HV    & CellViT              & 0.495                & 0.465              & 0.344                 & 0.335                 & 0.067             \\
          $\text{CellViT}_{256}-\text{Over}$                                                           & HV    & CellViT              & 0.494                & 0.467              & 0.349                 & 0.339                 & 0.071             \\
          $\text{CellViT}_{256}-\text{Aug}$                                                            & HV    & CellViT              & 0.565                & 0.558              & 0.419                 & 0.403                 & \textbf{0.156}    \\
          $\text{CellViT}_{256}-\text{No-FC}$                                                          & HV    & CellViT              & 0.567                & 0.548              & 0.416                 & 0.404                 & 0.141             \\ \midrule  

          $\text{CellViT}_{256}$                                                                       & SD    & CellViT              & 0.516                & 0.507              & 0.400                 & 0.331                 & 0.128             \\
          CellViT-SAM-H                                                                                & SD    & CellViT              & 0.548                & 0.544              & 0.400                 & 0.347                 & 0.132             \\
          $\text{CellViT}_{256}$                                                                       & CPP   & CellViT              & 0.540                & 0.524              & 0.414                 & 0.369                 & 0.133             \\ 
          CellViT-SAM-H                                                                                & CPP   & CellViT              & 0.571                & 0.565              & 0.405                 & 0.395                 & 0.131             \\ \midrule
          \makecell[l]{$\text{CellViT}_{256}$*** \\ ($0.50~\si{\micro\meter \per px}$)}                & HV    & CellViT              & 0.497                & 0.467              & 0.292                 & 0.285                 & 0.021             \\
          \makecell[l]{CellViT-SAM-H*** \\ ($0.50~\si{\micro\meter \per px}$)}                         & HV    & CellViT              & 0.528                & 0.502              & 0.315                 & 0.311                 & 0.031             \\ \bottomrule
          \end{tabular}
      }
  \end{table}
\begin{table*}[!t]
\centering
\caption{Average $mPQ$ and $bPQ$ across the 19 tissue types of the PanNuke dataset for three-fold cross-validation. The standard deviation (STD) of the splits is provided in the final row. STARDIST models with with ResNet50 (RN50) encoder were re-trained with CPP-Net hyperparameters (CPP-HP) and CellViT hyperparameters (CellViT-HP) for comparison. For the CellViT models, just the architecture with HoVer-Net decoder (HV-Net) is given.     
*TSFD-Net was not evaluated on the official three-fold splits of the PanNuke dataset and left out by the comparison **STARDIST trained by \citet{cpp_net} ***Model re-trained by ourselves}
\label{tab:tissue_seg_results}
\resizebox{\textwidth}{!}{%
\begin{tabular}{@{}llllllllllllllllllllllll@{}}
\toprule
     & \multicolumn{2}{l}{HoVer-Net}  &  & \multicolumn{2}{l}{TSFD-Net*} &  & \multicolumn{2}{l}{STARDIST**}  &  & \multicolumn{2}{l}{STARDIST***} &  & \multicolumn{2}{l}{STARDIST***}   &  & \multicolumn{2}{l}{CPP-Net}         &  & \multicolumn{2}{l}{$\text{CellViT}_{256}$} &  & \multicolumn{2}{l}{CellViT-SAM-H} \\ \cmidrule(lr){8-9} \cmidrule(lr){11-12} \cmidrule(lr){14-15} \cmidrule(lr){17-18} \cmidrule(lr){20-21} \cmidrule(lr){23-24}
     & \multicolumn{2}{l}{}  &  & \multicolumn{2}{l}{} &  & \multicolumn{2}{l}{\makecell[l]{\small RN50 encoder}}  &  & \multicolumn{2}{l}{\makecell[l]{\small RN50 encoder \\ \small CPP-HP}} &  & \multicolumn{2}{l}{\makecell[l]{\small RN50 encoder \\ \small CellViT-HP}}   &  & \multicolumn{2}{l}{\small RN50 encoder}         &  & \multicolumn{2}{l}{\makecell[l]{\small HV-Net decoder \\ \small CellViT-HP}} &  & \multicolumn{2}{l}{\makecell[l]{\small HV-Net decoder \\ \small CellViT-HP }}  \\ \midrule
           Tissue & $mPQ$           & $bPQ$            &  & $mPQ$           & $bPQ$           &  & $mPQ$           & $bPQ$           &  & $mPQ$           & $bPQ$                                  &  & $mPQ$           & $bPQ$                                &  & $mPQ$             & $bPQ$               &  & $mPQ$                         & $bPQ$          &  & $mPQ$             & $bPQ$              \\ \midrule
Adrenal    & 0.4812        & 0.6962         &  & 0.5223        & 0.6900        &  & 0.4868        & 0.6972        &  & 0.4928        & 0.6954                               &  & 0.4834        & 0.6884                             &  & 0.4922          & 0.7031            &  & 0.4950                      & 0.7009       &  & \textbf{0.5134}          & \textbf{0.7086}           \\
Bile Duct  & 0.4714        & 0.6696         &  & 0.5000        & 0.6284        &  & 0.4651        & 0.6690        &  & 0.4632        & 0.6583                               &  & 0.4680        & 0.6564                             &  & 0.4650          & 0.6739            &  & 0.4721                      & 0.6705       &  & \textbf{0.4887}          & \textbf{0.6784}           \\
Bladder    & 0.5792        & 0.7031         &  & 0.5738        & 0.6773        &  & 0.5793        & 0.6986        &  & 0.5643        & 0.6949                               &  & 0.5730        & 0.6901                             &  & \textbf{0.5932} & 0.7057            &  & 0.5756                      & 0.7056       &  & 0.5844          & \textbf{0.7068}           \\
Breast     & 0.4902        & 0.6470         &  & 0.5106        & 0.6245        &  & 0.5064        & 0.6666        &  & 0.4948        & 0.6585                               &  & 0.4889        & 0.6497                             &  & 0.5066          & 0.6718            &  & 0.5089                      & 0.6641       &  & \textbf{0.5180}          & \textbf{0.6748}           \\
Cervix     & 0.4438        & 0.6652         &  & 0.5204        & 0.6561        &  & 0.4628        & 0.6690        &  & 0.4752        & 0.6739                               &  & 0.4781        & 0.6685                             &  & 0.4779          & \textbf{0.6880}            &  & 0.4893                      & 0.6862       &  & \textbf{0.4984}          & 0.6872           \\
Colon      & 0.4095        & 0.5575         &  & 0.4382        & 0.5370        &  & 0.4205        & 0.5779        &  & 0.4230        & 0.5704                               &  & 0.4087        & 0.5555                             &  & 0.4269          & 0.5888            &  & 0.4245                      & 0.5700       &  & \textbf{0.4485}          & \textbf{0.5921}           \\
Esophagus  & 0.5085        & 0.6427         &  & 0.5438        & 0.6306        &  & 0.5331        & 0.6655        &  & 0.5200        & 0.6508                               &  & 0.5175        & 0.6446                             &  & 0.5410          & \textbf{0.6755}            &  & 0.5373                      & 0.6619       &  & \textbf{0.5454}          & 0.6682           \\
Head \& Neck & 0.4530        & 0.6331         &  & 0.4937        & 0.6277        &  & 0.4768        & 0.6433        &  & 0.4660        & 0.6305                               &  & 0.4629        & 0.6215                             &  & 0.4667          & 0.6468            &  & 0.4901                      & 0.6472       &  & \textbf{0.4913}          & \textbf{0.6544}           \\
Kidney     & 0.4424        & 0.6836         &  & 0.5517        & 0.6824        &  & \textbf{0.5880}        & 0.6998        &  & 0.5090        & 0.6888                               &  & 0.4750        & 0.6800                             &  & 0.5092          & 0.7001            &  & 0.5409                      & 0.6993       &  & 0.5366          & \textbf{0.7092}           \\
Liver      & 0.4974        & 0.7248         &  & 0.5079        & 0.6675        &  & 0.5145        & 0.7231        &  & 0.4899        & 0.7106                               &  & 0.5034        & 0.7051                             &  & 0.5099          & 0.7271            &  & 0.5065                      & 0.7160       &  & \textbf{0.5224}           & \textbf{0.7322}           \\
Lung       & 0.4004        & 0.6302         &  & 0.4274        & 0.5941        &  & 0.4128        & 0.6362        &  & 0.3627        & 0.6087                               &  & 0.3931        & 0.6205                             &  & 0.4234          & 0.6364            &  & 0.4102                      & 0.6317       &  & \textbf{0.4314}           & \textbf{0.6426}           \\
Ovarian    & 0.4863        & 0.6309         &  & 0.5253        & 0.6431        &  & 0.5205        & 0.6668        &  & 0.5106        & 0.6573                               &  & 0.5204        & 0.6547                             &  & 0.5276          & \textbf{0.6792}            &  & 0.5260                      & 0.6596       &  & \textbf{0.5390}           & 0.6722           \\
Pancreatic & 0.4600        & 0.6491         &  & 0.4893        & 0.6241        &  & 0.4585        & 0.6601        &  & 0.4548        & 0.6516                               &  & 0.4526        & 0.6439                             &  & 0.4680          & \textbf{0.6742}            &  & \textbf{0.4769}                       & 0.6643       &  & 0.4719          & 0.6658           \\
Prostate   & 0.5101        & 0.6615         &  & 0.5431        & 0.6406        &  & 0.5067        & 0.6748        &  & 0.4905        & 0.6561                               &  & 0.4812        & 0.6457                             &  & 0.5261          & \textbf{0.6903}            &  & 0.5164                      & 0.6695       &  & \textbf{0.5321}           & 0.6821           \\
Skin       & 0.3429        & 0.6234         &  & 0.4354        & 0.6074        &  & 0.3610        & 0.6289        &  & 0.3826        & 0.6349                               &  & 0.3709        & 0.6197                             &  & 0.3547          & 0.6192            &  & 0.3661                      & 0.6400       &  & \textbf{0.4339}           & \textbf{0.6565}           \\
Stomach    & \textbf{0.4726}        & 0.6886         &  & 0.4871        & 0.6529        &  & 0.4477        & 0.6944        &  & 0.4239        & 0.6769                               &  & 0.4194        & 0.6642                             &  & 0.4553          & \textbf{0.7043}            &  & 0.4475                      & 0.6918       &  & 0.4705           & 0.7022           \\
Testis     & 0.4754        & 0.6890         &  & 0.4843        & 0.6435        &  & 0.4942        & 0.6869        &  & 0.4819        & 0.6848                               &  & \textbf{0.5141}& 0.6812                             &  & 0.4917          & \textbf{0.7006}            &  & 0.5091                      & 0.6883       &  & 0.5127          & 0.6955           \\
Thyroid    & 0.4315        & 0.6983         &  & 0.5154        & 0.6692        &  & 0.4300        & 0.6962        &  & 0.4246        & 0.6962                               &  & 0.4175        & 0.6921                             &  & 0.4344          & 0.7094            &  & 0.4412                      & 0.7035       &  & \textbf{0.4519}           & \textbf{0.7151}           \\
Uterus     & 0.4393        & 0.6393         &  & 0.5068        & 0.6204        &  & 0.4480        & 0.6599        &  & 0.4452        & 0.6455                               &  & 0.4683        & 0.6428                             &  & \textbf{0.4790}          & 0.6622            &  & 0.4737                      & 0.6516       &  & 0.4737          & \textbf{0.6625}          \\ \midrule
Average    & 0.4629        & 0.6596         &  & 0.5040        & 0.6377        &  & 0.4796        & 0.6692        &  & 0.4671        & 0.6602                               &  & 0.4682        & 0.6539                             &  & 0.4815          & 0.6767            &  & 0.4846                      & 0.6696       &  & \textbf{0.4980}          & \textbf{0.6793}           \\
STD        & 0.0076        & 0.0036         &  & -             & -             &  & -             & -             &  & 0.0489        & 0.0340                               &  & 0.0496        & 0.0348                             &  & -               & -                 &  & 0.0503                      & 0.0340       &  & 0.0413          & 0.0318           \\ \bottomrule
\end{tabular}%
}
\end{table*}

\begin{figure*}[!t]
	\centering
	\includegraphics[]{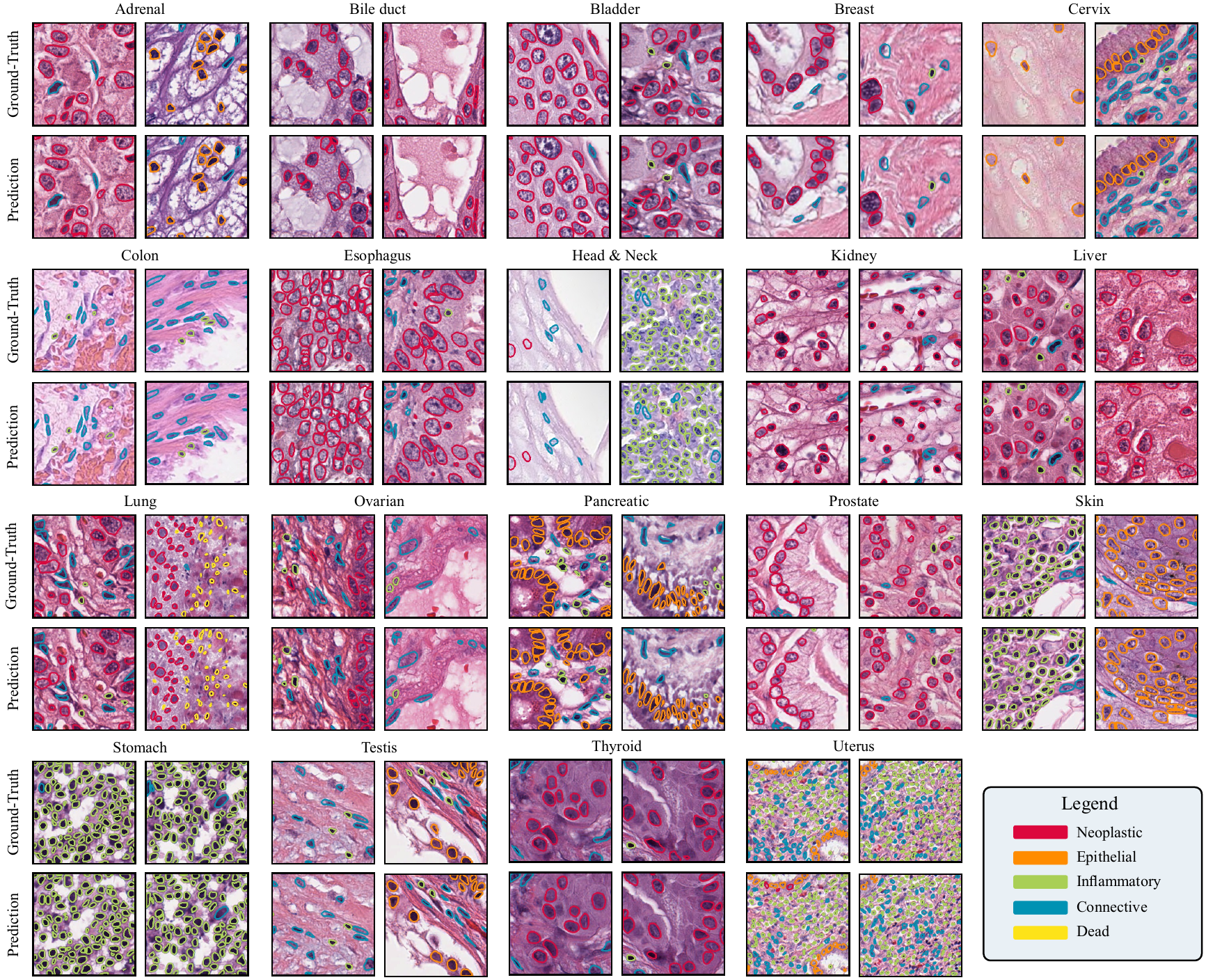}
	\caption{Example of PanNuke patches with ground-truth annotations and CellViT-SAM-H predictions overlaid for each tissue type.} 
	\label{fig:pannuke_example_image}
     \vspace{-3mm}
\end{figure*}

\begin{table*}[!t]
    \centering
    \caption{MoNuSeg validation result for $\text{CellViT}_{256}$ and CellViT-SAM-H models with HoVer-Net decoder and trained with CellViT hyperparameters on different dataset resolutions and inference patch sizes averaged over all three PanNuke training folds. The original image size for $0.25~\si{\micro\meter \per px}$ resolution with $\times 40$ magnification (mag.) is $1024~\text{px}$, and $512~\text{px}$ for $0.25~\si{\micro\meter \per px}$ ($\times 20$ mag.). *Models trained on downscaled $0.50~\si{\micro\meter \per px}$ PanNuke images}
    \label{tab:monuseg_results}
    \resizebox{\textwidth}{!}{%
        \begin{tabular}{llccccccccccccccc}
        \toprule
        \multirow{2}{*}{\begin{tabular}[c]{@{}l@{}}Dataset\\ resolution\end{tabular}}                             & Inference patch size &  & \multicolumn{4}{c}{$256~\text{px}$ with $64~\text{px}$ overlap}         &  & \multicolumn{4}{c}{$256~\text{px}$ without overlap}         &  & \multicolumn{4}{c}{$1024~\text{px}$ (no patching)}         \\ \cmidrule(lr){2-2} \cmidrule(lr){4-7} \cmidrule(lr){9-12} \cmidrule(l){14-17} 
                                                                        & Metric               &  & $bPQ$     & $P_\text{d}$     & $R_\text{d}$     & $F_1,\text{d}$     &  & $bPQ$     & $P_\text{d}$     & $R_\text{d}$     & $F_1,\text{d}$     &  & $bPQ$     & $P_\text{d}$     & $R_\text{d}$     & $F_1,\text{d}$     \\ \midrule
        \multirow{4}{*}{\begin{tabular}[c]{@{}l@{}}$0.25~\si{\micro\meter \per px}$ \\ ($\times 40$ mag.) \end{tabular}} & $\text{CellViT}_{256}$            &  & 0.660 & 0.841 & 0.886 & 0.863 &  & 0.621 & 0.814 & 0.897 & 0.853 &  & 0.661 & 0.838 & 0.859 & 0.848 \\
                                                                        & CellViT-SAM-H      &  & 0.671 & 0.846 & 0.893 & \textbf{0.868} &  & 0.631 & 0.814 & 0.906 & 0.857 &  & \cellcolor[rgb]{0.753,0.753,0.753}\textbf{0.672} & \cellcolor[rgb]{0.753,0.753,0.753}\textbf{0.847} & \cellcolor[rgb]{0.753,0.753,0.753} 0.885 & \cellcolor[rgb]{0.753,0.753,0.753} 0.865 \\
                                                                        & $\text{CellViT}_{256}$ ($0.50~\si{\micro\meter \per px}$)*             &  & 0.509 & 0.748 & 0.893 & 0.804 &  & 0.491 & 0.728 & 0.895 & 0.792 &  & 0.515 & 0.759 & 0.905 & 0.813 \\
                                                                        & CellViT-SAM-H ($0.50~\si{\micro\meter \per px}$)*               &  & 0.524 & 0.746 & 0.963 & 0.840 &  & 0.514 & 0.729 & 0.963 & 0.829 &  & 0.540 & 0.749 & \textbf{0.966} & 0.842 \\ \midrule
                                                                        &                      &  & \multicolumn{4}{c}{$256~\text{px}$ with $64~\text{px}$ overlap}         &  & \multicolumn{4}{c}{$256~\text{px}$ without overlap}         &  & \multicolumn{4}{c}{$512~\text{px}$ (no patching)}         \\ \midrule
        \multirow{4}{*}{\begin{tabular}[c]{@{}l@{}}$0.50~\si{\micro\meter \per px}$ \\ ($\times 20$ mag.)\end{tabular}} & $\text{CellViT}_{256}$              &  & 0.588 & 0.918 & 0.766 & 0.834 &  & 0.586 & 0.902 & 0.759 & 0.824 &  & 0.593 & 0.919 & 0.771 & 0.837 \\
                                                                        & CellViT-SAM-H         &  & 0.627 & \textbf{0.922} & 0.791 & \textbf{0.851} &  & 0.620 & 0.908 & 0.784 & 0.841 &  & 0.627 & 0.909 & 0.792 & 0.846 \\
                                                                        & $\text{CellViT}_{256}$  ($0.50~\si{\micro\meter \per px}$)*                  &  & 0.643 & 0.874 & 0.803 & 0.836 &  & 0.640 & 0.867 & 0.797 & 0.830 &  & 0.644 & 0.873 & 0.810 & 0.840 \\
                                                                        & CellViT-SAM-H ($0.50~\si{\micro\meter \per px}$)*               &  & 0.649 & 0.835 & 0.814 & 0.824 &  & 0.648 & 0.841 & 0.820 & 0.830 &  & \cellcolor[rgb]{0.753,0.753,0.753}\textbf{0.655} & \cellcolor[rgb]{0.753,0.753,0.753} 0.840 & \cellcolor[rgb]{0.753,0.753,0.753}\textbf{0.829} & \cellcolor[rgb]{0.753,0.753,0.753} 0.834 \\ \bottomrule
        \end{tabular}%
    }
\end{table*}

\begin{figure*}[!t]
    \centering
    \includegraphics{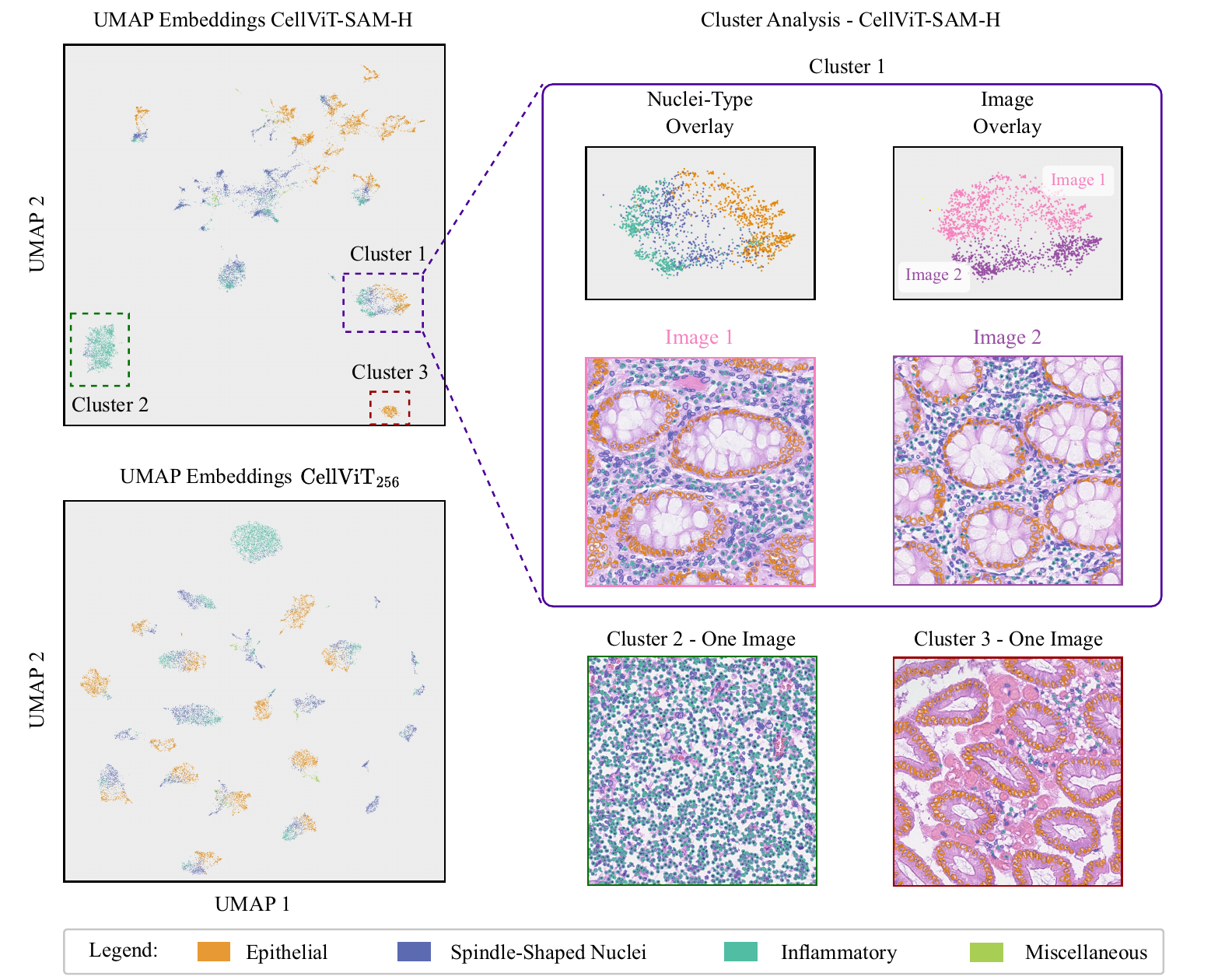}
    \caption{Two-dimensional UMAP embedding visualization (left) of the CoNSeP dataset with the CellViT-SAM-H and $\text{CellViT}_{256}$ (HoVer-Net encoder) models trained on PanNuke. We extract cell-tokens for each detected cell with our model, resulting in one embedding vector per cell. On the right side of the figure, representative clusters derived with the CellViT-SAM-H model are displayed alongside corresponding tissue images. The color overlay illustrates the ground-truth nuclei types within the dataset.}
    \label{fig:umap_cellvit}
    \vspace{-4mm}
\end{figure*}

\subsection{Segmentation Quality on PanNuke}
\label{subsec:comparison_with_other_algorithms}

To assess the segmentation quality, the panoptic quality is used. Tab. \ref{tab:pq_nuclei} presents the $PQ$ values for each nuclei type, averaged over all tissue types. Among all settings, CellViT$_{256}$ and CellViT-SAM-H networks with HoVer-Net decoder excell in neoplastic, connective, and epithelial nuclei. However, in the case of inflammatory and connective nuclei, they are outperformed by TSFD-Net due to its larger training dataset (80/20 split vs. 33/67 split). Notably, all models consistently yield the lowest results for dead cells, attributed to class imbalance and the small size of dead cells. 
To further analyze the influence of Focal Tversky loss and our custom oversampling strategy, we included $PQ$ values for the $\text{CellViT}_{256}$ model (HoVer-Net decoder) with different regularization techniques in Tab. \ref{tab:pq_nuclei}. It is observed that the segmentation quality is improved by oversampling ($\text{CellViT}_{256}\text{-Over}$) for almost all nuclei classes except neoplastic nuclei. The deterioration of neoplastic nuclei is attributed to the class rebalancing, as neoplastic nuclei constitute the majority class in the dataset. Removing the Focal Tversky loss ($\text{CellViT}_{256}\text{-No-FC}$), leads to a decrease in panoptic quality for all classes, except neoplastic nuclei again. Models employing STARDIST and CPP-Net decoders achieve lower panoptic quality than HoVer-Net decoder models but surpass baseline networks. The results for the $0.50~\si{\micro\meter \per px}$ models reveal a significant drop in performance when using the downscaled data. \\
Finally, we evaluate the segmentation performance of the CellViT$_{256}$ and CellViT-SAM-H models with HoVer-Net decoder against the best baseline models by computing the binary $PQ$ ($bPQ$) and the more challenging multi-class PQ ($mPQ$) for each of the 19 tissue types in PanNuke, providing an assessment of both instance segmentation qualities. As baseline experiments, we just include the best HoVer-Net model by \citet{Hovernet}, TSFD-Net and the original STARDIST and CPP-Net models with ResNet50 encoder \citet{cpp_net}. For our detection experiments in Sec. \ref{subsec:ablation_study}, we retrained the baseline STARDIST model with the ResNet50 encoder. Even though we are not able to reproduce segmentation results reported by \citet{cpp_net} with CPP-Net and our hyperparameter settings, we include all three results in Tab. \ref{tab:tissue_seg_results} for a fair comparison (STARDIST ResNet50 \citep{cpp_net}, STARDIST ResNet50 re-trained with CPP-Net hyperparameters, and STARDIST ResNet50 with our hyperparameters).
Our experimental results demonstrate that CPP-Net and STARDIST (both with ResNet50 encoder) exhibit comparable $bPQ$ values, whereas our CellViT models achieve superior $mPQ$. This is primarily attributed to the superior detection capabilities of our models, which significantly impacts the $mPQ$ value.
The best average model is CellViT-SAM-H with the HoVer-Net decoder architecture trained with our hyperparameter settings. Segmentation results per tissue for $0.50~\si{\micro\meter \per px}$ are given in the Appendix \ref{tab:tissue_seg_results_appendix}. \\
To provide a visual representation of the segmentations, we include tissue-wise comparisons between ground-truth and segmentation predictions of the CellViT-SAM-H model in Fig. \ref{fig:pannuke_example_image}. As observed in the lung example, the instance segmentation of dead cells poses a significant challenge due to their small size. Furthermore, detecting and segmenting dead nuclei becomes even more difficult when these images are scaled down from $0.25~\si{\micro\meter \per px}$ to $0.50~\si{\micro\meter \per px}$ resolution. 

\subsection{MoNuSeg Test Performance}
\label{subsec:generalization_performance}

In this experiment, we focused on instance segmentation without classification on the MoNuSeg dataset to assess the generalizability of our models (just with HoVer-Net decoder) at resolutions of $0.25~\si{\micro\meter \per px}$ and $0.50~\si{\micro\meter \per px}$. Additionally, we aim to evaluate the impact of changing the input sequence size by performing inference on large-scale tiles of size $1024~\text{px}$ ($0.25~\si{\micro\meter \per px}$) and $512~\text{px}$ ($0.50~\si{\micro\meter \per px}$), respectively, comparing the results to non-overlapping $256~\text{px}$ patches and $256~\text{px}$ patches with an overlap of $64~\text{px}$ derived by a shifting window approach. We utilized the three final models of the PanNuke training folds for each architecture and conducted inference on the MoNuSeg data without retraining. 
The evaluation results are presented in Tab. \ref{tab:monuseg_results}. Consistent with the previous experiments, the CellViT-SAM-H model is the best-performing model. It achieves a $bPQ$-score of $0.672$ on $1024~\text{px}$ tiles when no patching was applied and of $0.671$ for $256~\text{px}$ tiles with an overlap of $64~\text{px}$. However, when using $256~\text{px}$ patches without overlap, the $bPQ$-score decreases to $0.631$, likely due to the absence of merging overlapping nuclei at cell borders and double detected cells (higher recall). Importantly, the overall comparison between larger tiles and smaller tiles with overlapping indicates that inference on larger tiles did not lead to a degradation in performance. This justifies our inference pipeline for large-scale WSI, in which we are using $1024~\text{px}$ sized patches with an overlap of $64~\text{px}$ and overlapping merging strategies. The $\text{CellViT}_{256}$ model yields slightly inferior results compared to the CellViT-SAM-H model. \\
Using the models trained on $0.50~\si{\micro\meter \per px}$ data on the $0.25~\si{\micro\meter \per px}$ data and vice versa, the $0.50~\si{\micro\meter \per px}$ trained models exhibit poor performance on $0.25~\si{\micro\meter \per px}$ data, while the $0.25~\si{\micro\meter \per px}$ trained models experience a less severe performance drop on the $0.50~\si{\micro\meter \per px}$ data. Nevertheless, networks trained and evaluated on the same WSI resolution achieved the best performance, thus it is advisable to align image resolution between different dataset and use the appropriate model. Consistently, the best results are achieved for WSI acquired with a resolution of $0.25~\si{\micro\meter \per px}$. \\
We include a visual demonstration presenting a tissue tile from the MoNuSeg test set along with binary segmentation masks generated by the CellViT-SAM-H and CellViT-SAM-H ($0.50~\si{\micro\meter \per px}$) models in the Appendix.

\subsection{Token Analysis}
\label{subsec:token_analysis}
In Figure \ref{fig:umap_cellvit}, we present the two-dimensional UMAP embeddings of cell tokens from the CoNSeP dataset. The CellViT-SAM-H and $\text{CellViT}_{256}$ models with HoVer-Net decoder, trained on the PanNuke dataset, were utilized. The tokens were extracted simultaneously with cell detections in a single inference pass. The color overlay in the scatter plots (left) and tissue images (right) indicates the respective nuclei classes. Consistent with \citet{Hovernet}, we grouped normal and malignant/dysplastic epithelial nuclei into an "epithelial" class, while fibroblast, muscle, and endothelial nuclei were grouped into the "spindle-shaped nuclei" class. The global clusters in the scatter plot represent cells from different images, with clusters containing cells from the same tissue phenotype being grouped together.
An example of this is cluster 1 for the CellViT-SAM-H model. It comprises cell clusters from two images, both containing multiple glands. Within this cluster, the local spatial arrangement of the cell embeddings allows differentiation of nuclei types (epithelial, spindle-shaped, and inflammatory) despite the model not being explicitly trained for all cell classes (spindle-shaped cells are not explicitly defined in the PanNuke dataset). Cluster 3, which is spatially close to cluster 1, contains even more glands, while the tissue image associated with the distant cluster 2 lacks glands and primarily consists of spindle-shaped and inflammatory nuclei. In summary, the global UMAP arrangement primarily captures differences in the nuclei's tissue environment (e.g., nearby glands, muscles). The local arrangement highlights distinctions between nuclei without the need for fine-tuning the model for specific nuclei types. Notably, for the $\text{CellViT}_{256}$ model, the global tissue differences are even more pronounced. To quantitatively assess the quality of the embeddings, we trained a linear nuclei classifier (Appendix) on the embeddings of the training data (15,548 nuclei) to classify the nuclei into the CoNSeP classes. We evaluated the classifier on the embeddings of the test images (8,773 nuclei). The model achieved an area under the receiver operating characteristics curve (AUROC) of 0.963 for the validation data using the CellViT-SAM-H embeddings. When utilizing the $\text{CellViT}_{256}$ embeddings, the model achieved an AUROC of 0.960. This demonstrates the effectiveness of our embeddings in classifying unknown nuclei classes, with both CellViT-SAM-H and $\text{CellViT}_{256}$ embeddings yielding high AUROC values.

\subsection{Inference Runtime}
\label{subsec:inference_runtime}
Our inference runtime benchmark shows that our inference pipeline is accelerated by a factor of $2.49$ ($\text{CellViT}_{256}$) and $2.25$ (CellViT-SAM-H) when using $1024~\text{px}$ input patches instead of $256~\text{px}$. The $\text{CellViT}_{256}$ model with $1024~\text{px}$ input patches is 1.34 times faster than the CellViT-SAM-H model with $1024~\text{px}$ patches. Both CellViT models with our large $1024~\text{px}$ input patch size outperform the HoVerNet model, with speedups of $1.85$ ($\text{CellViT}_{256}$) and $1.39$ (CellViT-SAM-H), respectively.

\section{Discussion and Conclusion}
\label{sec:discussion_and_conclusion}
Nuclei instance segmentation is crucial for clinical applications, requiring automated tools that offer high robustness and reliability. In the clinical context of performing large-scale analysis on clinical patient cohorts, accurate detection is considered more important than precise segmentation. \\
In this work, we introduced a novel deep learning-based method for simultaneously segmenting and detecting nuclei in digitized H\&E tissue samples. Our work was inspired by the success of previous works using large-scale trained Vision Transformers, particularly by the contributions made by \citet{vit256} ($\text{ViT}_{256}$) and \citet{SAM} (SAM). The CellViT network proposed in this study demonstrates state-of-the-art performance for both nuclei instance segmentation and nuclei detection on the PanNuke dataset. Additionally, the results on the MoNuSeg dataset validate the generalizability of our model to previously unseen cohorts. Notably, our model surpasses all other existing methods by a significant margin for nuclei detection and classification, elevating nuclei detection in H\&E-slides to a new level. By leveraging the most recent computer vision approaches, we showed that both in-domain pre-training ($\text{ViT}_{256}$) and the use of the SAM foundation model yields significantly better results compared to randomly initialized network weights. Our larger inference patch size allows us to be $1.85$ times faster than the popular HoVer-Net inference framework by \citet{Hovernet}, which could save hours in computational time when dealing with huge gigapixel WSI. Moreover, our framework allows direct assessment of a localizable ViT-token from a detected nucleus that can be further used in downstream tissue analysis tasks. Although an evaluation of this aspect is pending, we anticipate promising prospects based on our first results in Sec. \ref{subsec:token_analysis}. Our work provides the potential to design interpretable algorithms that directly correlate with specific cells or cell patterns. One possible direction for future research involves graph-based networks with attention mechanisms using these embeddings. \\
Nevertheless, external validation of the results is necessary. Yet, additional datasets are required, especially to verify the detection quality of our model. Furthermore, our models exhibit reliable performance only for WSI acquired at $0.25~\si{\micro\meter \per px}$ resolution.
While the results obtained with $0.50~\si{\micro\meter \per px}$ images are acceptable in terms of detection, there is room for improvement, as there is a huge performance gap between $0.25~\si{\micro\meter \per px}$ and  $0.50~\si{\micro\meter \per px}$-WSI processing. We recommend
to scan the tissue samples on a resolution of  $0.25~\si{\micro\meter \per px}$ if technically possible.
In the future, we plan to apply the proposed model with extracted nuclei tokens to downstream histological image analysis tasks. This will enable us to validate if simultaneously extracted tokens are an advantage for building interpretable algorithms for computational pathology. Additionally, it will allow us to evaluate which tokens have achieved a more meaningful representation of the tissue and are better suited for downstream tasks, as there are just minimal differences in the segmentation and detection performance between our best-performing $\text{CellViT}_{256}$ and CellViT-SAM-H models. To ensure the accessibility of our results, we have made both the code and pre-trained models publicly available under an open-source license for non-commercial use.
\vspace{+2mm}

\section*{CRediT authorship contribution statement} 
\textbf{Fabian Hörst:} Conceptualization, Methodology, Software, Formal Analysis, Investigation, Data Curation, Writing - Original Draft, Writing - Review \& Editing, Visualization.
\textbf{Moritz Rempe:} Methodology, Writing - Original Draft, Writing - Review \& Editing.
\textbf{Lukas Heine:} Methodology, Writing - Original Draft, Writing - Review \& Editing.
\textbf{Constantin Seibold:} Conceptualization, Writing - Review \& Editing.
\textbf{Julius Keyl:} Validation, Writing - Review \& Editing.
\textbf{Giulia Baldini:} Validation, Writing - Review \& Editing.
\textbf{Selma Ugurel:} Validation, Writing - Review \& Editing.
\textbf{Jens Siveke} Validation, Writing - Review \& Editing.
\textbf{Barbara Grünwald:} Validation, Writing - Review \& Editing.
\textbf{Jan Egger:} Writing - Review \& Editing, Supervision.
\textbf{Jens Kleesiek:} Resources, Writing - Review \& Editing, Supervision, Project administration, Funding acquisition.

\section*{Declaration of competing interest}

\textbf{S. U.} declares research support from Bristol Myers Squibb and Merck Serono; speakers and advisory board honoraria from Bristol Myers Squibb, Merck Sharp \& Dohme, Merck Serono, and Novartis; and meeting and travel support from Almirall, Bristol-Myers Squibb, IGEA Clinical Biophysics, Merck Sharp \& Dohme, Novartis, Pierre Fabre, and Sun Pharma; outside the submitted work. \\
\textbf{J.T.S.} receives honoraria as consultant or for continuing medical education presentations from AstraZeneca, Bayer, Boehringer Ingelheim, Bristol-Myers Squibb, Immunocore, MSD Sharp Dohme, Novartis, Roche/Genentech, and Servier. His institution receives research funding from Abalos Therapeutics, Boehringer Ingelheim, Bristol-Myers Squibb, Celgene, Eisbach Bio, and Roche/Genentech; he holds ownership and serves on the Board of Directors of Pharma15, all outside the submitted work. \\
All other authors declare that they have no known competing financial interests or personal relationships that could have appeared to influence the work reported in this paper.

\section*{Data availability} 
The used datasets are publicly available. All models and source-code are available online here: \url{https://github.com/TIO-IKIM/CellViT}

\section*{Acknowledgments}
This work received funding from \lq KITE' (Plattform für KI-Translation Essen) from the REACT-EU initiative (\url{https://kite.ikim.nrw/}, EFRE-0801977) and the Cancer Research Center Cologne Essen (CCCE).

\bibliography{bibliography}

\clearpage
\section*{Supplementary Material}
\onecolumn
\renewcommand\thefigure{A.\arabic{figure}}    
\setcounter{figure}{0}  
\renewcommand{\thetable}{A.\arabic{table}} 
\setcounter{table}{0}
\renewcommand{\thesection}{A.\arabic{section}} 
\setcounter{section}{0}

\section{STARDIST and CPP-Net}
\label{app:stardist_cpp_loss}
Due to the new probability branch $PD$ and the new radial distances branch $RD$, the loss function of STARDIST changes to: 
\begin{align}
    \label{eq:stardist_loss_fn}
    \mathcal{L}_\text{total} = \mathcal{L}_\text{PD} + \mathcal{L}_\text{RD} + \mathcal{L}_\text{NT}
\end{align}
with the individual loss branches
\begin{align}
    \begin{split}
        \mathcal{L}_\text{PD} &= \lambda_{\text{PD}_\text{BCE}}\mathcal{L}_\text{BCE} \\ 
        \mathcal{L}_\text{SD} &= \lambda_{\text{SD}_\text{MSE}}\mathcal{L}_\text{MSE}  \\
        \mathcal{L}_\text{NT} &= \lambda_{\text{NT}_\text{DICE}}\mathcal{L}_\text{DICE} + \lambda_{\text{NT}_\text{BCE}}\mathcal{L}_\text{BCE}
    \end{split}
\end{align}
and weighting factors $\lambda_{\text{PD}_{\text{BCE}}} = \lambda_{\text{SD}_{\text{MSE}}} = \lambda_{\text{NT}_{\text{DICE}}} = \lambda_{\text{NT}_{\text{BCE}}} = 1$.
For $\mathcal{L}_\text{PD}$ and $\mathcal{L}_\text{PD}$, the loss is weighted by the ground-truth object probabilities.
When using the CPP-Net networks, we used the same loss function as in eq. \eqref{eq:stardist_loss_fn}, but changed the nuclei type loss to $\mathcal{L}_\text{NT}=\lambda_{\text{NT}_\text{FT}}\mathcal{L}_\text{FT} + \lambda_{\text{NT}_\text{DICE}}\mathcal{L}_\text{DICE} + \lambda_{\text{NT}_\text{BCE}}\mathcal{L}_\text{BCE}$, with $\lambda_{\text{NT}_\text{FT}} = 0.5$, $\lambda_{\text{NT}_\text{DICE}}=0.2$ and $\lambda_{\text{NT}_\text{BCE}}=0.5$, as we achieved superior results with this setting.

\section{Supplementary Tables}

\begin{table}[H]
    \begin{minipage}{.475\linewidth}
    \centering
    \caption{Average $mPQ$ and $bPQ$ across the 19 tissue types of the PanNuke dataset for three-fold cross-validation for models trained on downscaled $0.50~\si{\micro\meter \per px}$ PanNuke images. The standard deviation (STD) of the splits is provided in the final row. Just the CellViT architecture with HoVer-Net decoder (HV-Net) is given. For comparison, we also included the networks trained and evaluated on original $0.25~\si{\micro\meter \per px}$ PanNuke images in the first two columns.
    *Models trained on downscaled $0.50~\si{\micro\meter \per px}$ PanNuke images}
    \label{tab:tissue_seg_results_appendix}
    \resizebox{\textwidth}{!}{%
    \begin{tabular}{@{}llllllllllllllllllllllll@{}}
    \toprule
         & \multicolumn{2}{l}{$\text{CellViT}_{256}$} &  & \multicolumn{2}{l}{CellViT-SAM-H} & & \multicolumn{2}{l}{$\text{CellViT}_{256}$*} & & \multicolumn{2}{l}{CellViT-SAM-H*} \\ \cmidrule(lr){2-3} \cmidrule(lr){5-6} \cmidrule(lr){8-9} \cmidrule(lr){11-12} 
         & \multicolumn{2}{l}{\makecell[l]{\small HV-Net decoder \\ \small CellViT-HP}} &  & \multicolumn{2}{l}{\makecell[l]{\small HV-Net decoder \\ \small CellViT-HP }} &  & \multicolumn{2}{l}{\makecell[l]{\small HV-Net decoder \\ \small CellViT-HP \\ $0.50~\si{\micro\meter \per px}$}} &  & \multicolumn{2}{l}{\makecell[l]{\small HV-Net decoder \\ \small CellViT-HP \\ $0.50~\si{\micro\meter \per px}$ }}  \\ \midrule
               Tissue & $mPQ$           & $bPQ$            &  & $mPQ$           & $bPQ$           &  & $mPQ$  & $bPQ$                 &  & $mPQ$    & $bPQ$      \\ \midrule
    Adrenal           & 0.4950          & 0.7009           &  & 0.5134          & 0.7086          &  & 0.3947 &	0.5967                &  & 0.4226	& 0.6139     \\
    Bile Duct         & 0.4721          & 0.6705           &  & 0.4887          & 0.6784          &  & 0.3594 &	0.5278                &  & 0.3791	& 0.5587     \\
    Bladder           & 0.5756          & 0.7056           &  & 0.5844          & 0.7068          &  & 0.3205 &	0.4221                &  & 0.3423	& 0.4457     \\
    Breast            & 0.5089          & 0.6641           &  & 0.5180          & 0.6748          &  & 0.4260 &	0.5761                &  & 0.4592	& 0.6097     \\
    Cervix            & 0.4893          & 0.6862           &  & 0.4984          & 0.6872          &  & 0.3713 &	0.5302                &  & 0.3967	& 0.5618     \\
    Colon             & 0.4245          & 0.5700           &  & 0.4485          & 0.5921          &  & 0.3139 &	0.4352                &  & 0.3485	& 0.4680     \\
    Esophagus         & 0.5373          & 0.6619           &  & 0.5454          & 0.6682          &  & 0.4485 &	0.5604                &  & 0.4574	& 0.5793     \\
    Head \& Neck      & 0.4901          & 0.6472           &  & 0.4913          & 0.6544          &  & 0.2597 &	0.3954                &  & 0.2821	& 0.4136     \\
    Kidney            & 0.5409          & 0.6993           &  & 0.5366          & 0.7092          &  & 0.3517 &	0.4805                &  & 0.3831	& 0.5203     \\
    Liver             & 0.5065          & 0.7160           &  & 0.5224          & 0.7322          &  & 0.3634 &	0.5415                &  & 0.3673	& 0.5659     \\
    Lung              & 0.4102          & 0.6317           &  & 0.4314          & 0.6426          &  & 0.3040 &	0.4261                &  & 0.3161	& 0.4489     \\
    Ovarian           & 0.5260          & 0.6596           &  & 0.5390          & 0.6722          &  & 0.4454 &	0.5691                &  & 0.4714	& 0.6033     \\
    Pancreatic        & 0.4769          & 0.6643           &  & 0.4719          & 0.6658          &  & 0.3395 &	0.4914                &  & 0.3465	& 0.5194     \\
    Prostate          & 0.5164          & 0.6695           &  & 0.5321          & 0.6821          &  & 0.3764 &	0.5243                &  & 0.3999	& 0.5404     \\
    Skin              & 0.3661          & 0.6400           &  & 0.4339          & 0.6565          &  & 0.2552 &	0.4481                &  & 0.2948	& 0.4835     \\
    Stomach           & 0.4475          & 0.6918           &  & 0.4705          & 0.7022          &  & 0.2948 &	0.5029                &  & 0.3105	& 0.5259     \\
    Testis            & 0.5091          & 0.6883           &  & 0.5127          & 0.6955          &  & 0.3856 &	0.5307                &  & 0.4031	& 0.5771     \\
    Thyroid           & 0.4412          & 0.7035           &  & 0.4519          & 0.7151          &  & 0.3527 &	0.6090                &  & 0.3758	& 0.6209     \\
    Uterus            & 0.4737          & 0.6516           &  & 0.4737          & 0.6625          &  & 0.3615 &	0.4972                &  & 0.3783	& 0.5384     \\ \midrule
    Average           & 0.4846          & 0.6696           &  & 0.4980          & 0.6793          &  & 0.3539 & 0.5087                &  & 0.3755   & 0.5366     \\
    STD               & 0.0503          & 0.0340           &  & 0.0413          & 0.0318          &  & 0.0546 & 0.0618                &  & 0.0541   & 0.0613     \\ \bottomrule
    \end{tabular}
    }
    \end{minipage}
    \begin{minipage}{0.05\linewidth}
        \hfill
    \end{minipage}
    \begin{minipage}{.475\linewidth}
    \centering
    \caption{Selected data augmentation techniques with probability and additional hyperparameters. Data augmentation is implemented with Albumentations. STARDIST just uses spatial transformations. For CPP-Net we used the same augmentations as for HoVer-Net decoder, because they achieved better results.}
    \label{tab:data_aug_hp}
    \resizebox{0.95\columnwidth}{!}{%
    \begin{tabular}{@{}lcl@{}}
    \toprule
    Augmentation Technique        & \multicolumn{1}{l}{Probability} & Hyperparamter                                                                                                     \\ \midrule
    90-degree rotation            & 0.5                             &                                                                                                                   \\
    Horizontal flipping           & 0.5                             &                                                                                                                   \\
    Vertical flipping             & 0.5                             &                                                                                                                   \\
    Downscaling                   & 0.15                            & \begin{tabular}[c]{@{}l@{}}max-scale: 0.5\\ min-scale: 0.5\end{tabular}                                           \\
    Blurring                      & 0.2                             & blur-limit: 10                                                                                                    \\
    Gaussian noise                & 0.25                            & var\_limit: 50                                                                                                    \\
    Color jittering               & 0.2                             & \begin{tabular}[c]{@{}l@{}}brightness: 0.25\\ contrast: 0.25\\ saturation: 0.1\\ hue: 0.05\end{tabular}           \\
    Superpixel representation     & 0.1                             & \begin{tabular}[c]{@{}l@{}}p\_replace: 0.1\\ n\_segments: 200\\ max-size: $H$/2\end{tabular}                      \\
    Zoom blur                     & 0.1                             & max-factor: 1.05                                                                                                  \\
    Random cropping with resizing & 0.1                             & crop-level: 0.5-1.0 of input size                                                                                 \\
    Elastic transformation        & 0.2                             & \begin{tabular}[c]{@{}l@{}}sigma: 25\\ alpha: 0.5\\ alpha-affine: 15\end{tabular}                                 \\
    Normalization                 & 1.0                             & \begin{tabular}[c]{@{}l@{}}Mean: $\left[0.5, 0.5, 0.5 \right]$\\ STD:  $\left[0.5, 0.5, 0.5 \right]$\end{tabular} \\ \bottomrule
    \end{tabular}%
    }
    \end{minipage}
\end{table}

\renewcommand{\arraystretch}{1.2}
\begin{table}[!h]
\centering
\caption{CellViT Hyperparameters for all training runs on the PanNuke dataset}
\label{tab:appendix-hyper}
\resizebox{\textwidth}{!}{%
\begin{tabular}{l|l}
\hline
Parameter & Value \\ \hline
Loss      & \begin{tabular}[c]{@{}l@{}}$\lambda_{\text{NP}_{\text{FT}}} = 1$, $\lambda_{\text{NP}_{\text{FT}}} = 1$, $\lambda_{\text{NP}_{\text{DICE}}} = 1$, $\lambda_{\text{HV}_{\text{MSE}}} = 2.5$, $\lambda_{\text{HV}_{\text{MSGE}}} = 8$, $\lambda_{\text{NT}_{\text{FT}}} = 0.5$, $\lambda_{\text{NT}_{\text{DICE}}} = 0.2$, $\lambda_{\text{NT}_{\text{BCE}}} = 0.5$, $\lambda_{\text{TC}_{\text{CE}}} = 0.1$,\\ $\alpha_\text{FT} = 0.7$, $\beta_\text{FT} = 0.3$, $\gamma_\text{FT} = 4/3$, $\varepsilon_\text{FT} = 1 \cdot 10^{-6}$ \end{tabular}       \\
Sampling  &  $\gamma_\text{s} = 0.85$     \\
Optimizer & AdamW \citep{adamw} \\
Training  &   $\eta = 3\cdot 10^{-4}$, $\lambda=1\cdot 10^{-4}$, $\beta_1 = 0.85$, $\beta_2 = 0.85$, $\text{epochs} = 130$, $\text{batch-size} = 16$, $\text{lr-scheduling} = 0.85$ \\ \hline
\end{tabular}%
}
\end{table}

\begin{table}[!h]
\centering
\caption{CPP-Net Hyperparameters for all training runs on the PanNuke dataset}
\label{tab:appendix-hyper}
\resizebox{\textwidth}{!}{%
\begin{tabular}{l|l}
\hline
Parameter & Value \\ \hline
Loss      & See Sec. \ref{app:stardist_cpp_loss}       \\
Sampling  &  $\gamma_\text{s} = 0.0$     \\
Optimizer & Adam \\
Training  &   $\eta = 3\cdot 10^{-4}$, $\lambda=1\cdot 10^{-3}$, $\beta_1 = 0.9$, $\beta_2 = 0.999$, $\text{epochs} = 130$, $\text{batch-size} = 16$, $\text{lr-scheduling} = \text{reducelronplateau}$ (multiply by 0.5) \\ \hline
\end{tabular}%
}
\end{table}

\clearpage

\section{Supplementary Figures}
\begin{figure*}[!h]
	\centering
	\includegraphics[width=\textwidth]{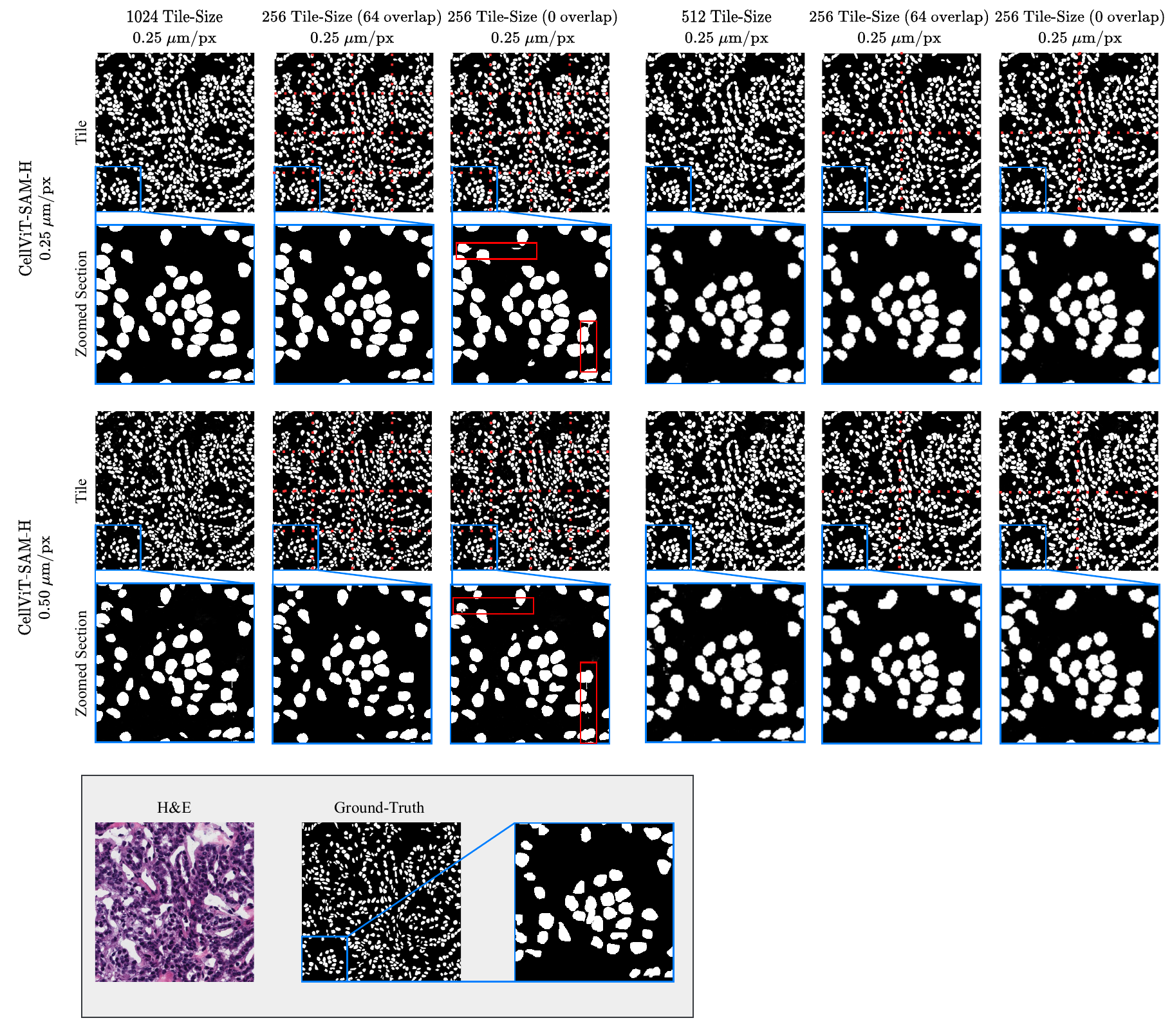}
	\caption{Example of one MoNuSeg tissue sample with ground-truth binary masks and predictions of the CellViT-SAM-H model for different input sizes and magnifications.} 
	\label{fig:monuseg_example_image}
\end{figure*}

\begin{figure}[!h]
    \centering
    \includegraphics{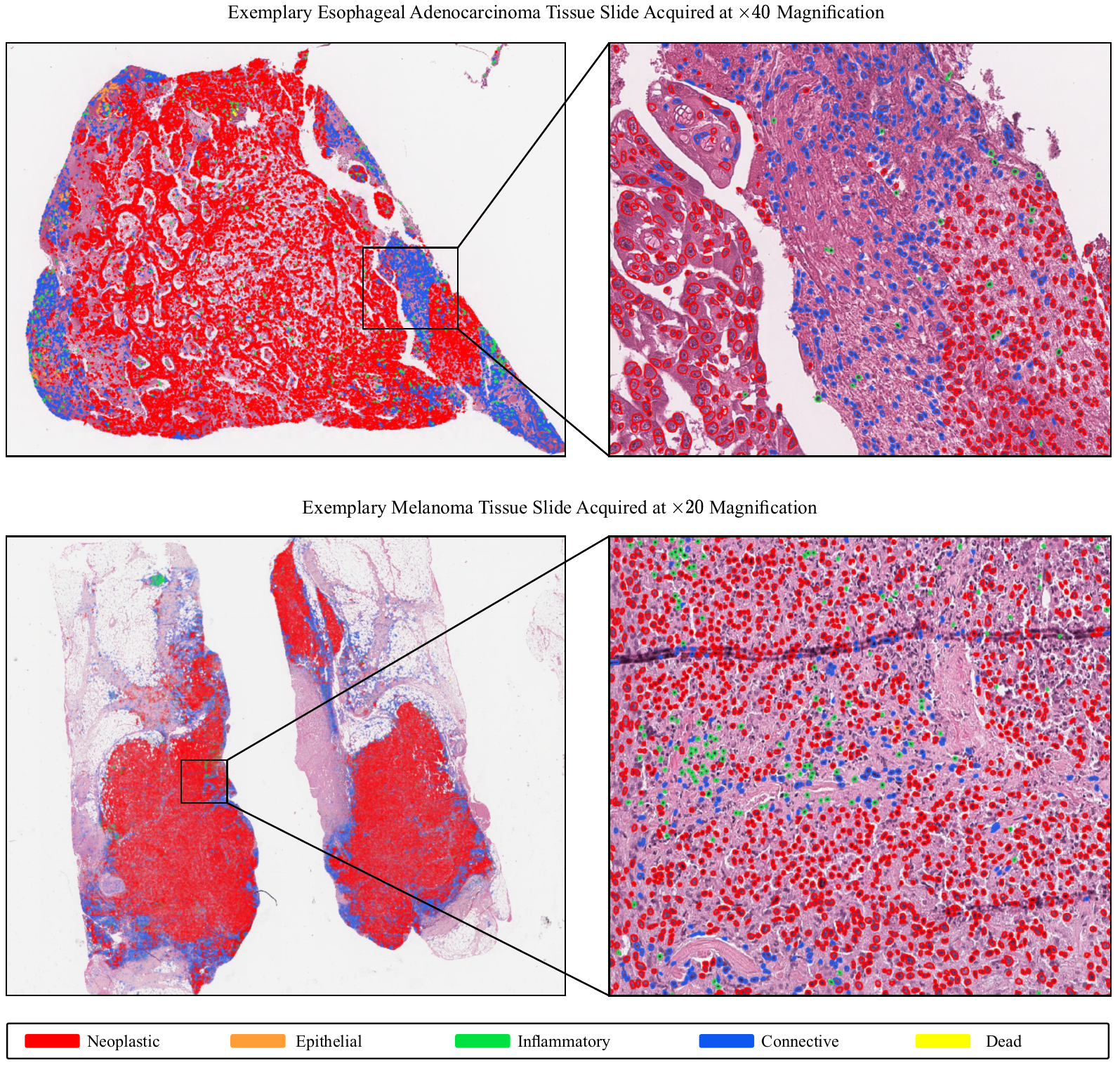}
    \caption{Exemplary WSI files with corresponding cell polygons imported into QuPath to show the interoperability of our inference pipeline. For each of the files, approximately 150,000 nuclei have been detected, which can be imported into QuPath without any performance problems regarding fast file loading and zooming on a standard laptop. The first WSI file was acquired at a magnification of $\times 40$ with $0.25~\si{\micro\meter \per px}$, the second at $\times 20$ with $0.50~\si{\micro\meter \per px}$.}
    \label{fig:qupath_import}
\end{figure}

\end{document}